\documentclass[acmsmall,screen,nonacm]{acmart}

\usepackage{preamble}
\usepackage{etoolbox}
\usepackage{microtype}
\usepackage{bussproofs}
\usepackage{dsfont}
\usepackage{bbm}
\usepackage{listings}
\usepackage{hhline}
\usepackage{caption}
\usepackage{subcaption}
\usepackage{mathtools}
\usepackage{booktabs}
\usepackage{tabularx}
\usepackage{multirow}
\usepackage{rotating}
\usepackage{makecell}
\usepackage{xcolor}
\usepackage{enumitem}
\usepackage{pgf}
\usepackage{float}
\usepackage{wrapfig}
\usepackage{listing}
\usepackage{listings-rust}
\usepackage{listings-verus}

\newcommand{\tr}[1]{}                 %
\newcommand{\eat}[1]{}                %

\newcommand{\ProtocolFour}[4]%
{\begin{eqnarray*}\mathsf{Protocol:}&&#1\\&&#2\\&&#3\\&&#4\end{eqnarray*}}

\def\centerhack#1{\hbox to 0pt{\hss\footnotesize #1\hss}}
\def\dchack#1{\vbox to 0pt{\vss{\hbox to 0pt{\hss#1\hss}}\vss}}

\def\FullBox{\hbox{\vrule width 8pt height 8pt depth 0pt}}
\def\qed{\ifmmode\qquad\FullBox\else{\unskip\nobreak\hfil
\penalty50\hskip1em\null\nobreak\hfil\FullBox
\parfillskip=0pt\finalhyphendemerits=0\endgraf}\fi}

\makeatletter
\def\sectionautorefnamesymbol{\S\@gobble}

\makeatother

\newcommand{\goalref}[1]{\hyperref[#1]{\textbf{G\ref*{#1}}}}
\newcommand{\chalref}[1]{\hyperref[#1]{\textbf{C\ref*{#1}}}}

\lstdefinestyle{DefaultStyle}
{
  basicstyle=\scriptsize\ttfamily\linespread{0.8}\selectfont,
  breaklines=true,
  tabsize=1,
  breakindent=2em,
  literate={\ \ }{{\ }}1
}

\lstdefinestyle{Verus}%
{ style=DefaultStyle
, basicstyle=\ttfamily\normalsize
, identifierstyle=%
, commentstyle=\color[gray]{0.4}%
, stringstyle=\color[rgb]{0, 0, 0.5}%
, keywordstyle={[1]\bfseries}%
, keywordstyle={[2]\color[rgb]{0.75, 0, 0}}%
, keywordstyle={[3]\color[rgb]{0, 0.5, 0}}%
, keywordstyle={[4]\color[rgb]{0, 0.5, 0}}%
, keywordstyle={[5]\color[rgb]{0, 0, 0.75}}%
, keywordstyle={[6]\color[rgb]{0, 0, 0.75}}%
, columns=spaceflexible%
, keepspaces=true%
, showspaces=false%
, showtabs=false%
, showstringspaces=true%
}%

\lstdefinestyle{VerusLineNos}{
  style=Verus%
, basicstyle=\ttfamily\scriptsize
, numbers=left%
, firstnumber=auto%
, numberblanklines=true%
, numberstyle=\color{gray}%
, numbersep=5pt%
, xleftmargin=15pt%
}

\lstdefinestyle{VerusLineNosCrStrike}{
  style=VerusLineNos%
, basicstyle=\scriptsize\ttfamily\linespread{0.8}\selectfont\color{red}%
, commentstyle=\color{red}%
, keywordstyle={[1]\color{red}}%
, keywordstyle={[2]\color{red}}%
, keywordstyle={[3]\color{red}}%
, keywordstyle={[4]\color{red}}%
, keywordstyle={[5]\color{red}}%
, keywordstyle={[6]\color{red}}%
, moredelim=[s][{\itshape\color{red}}]{\#[}{]}%
}

\newcolumntype{L}[1]{>{\raggedright\let\newline\\\arraybackslash\hspace{0pt}}m{#1}}
\newcolumntype{C}[1]{>{\centering\let\newline\\\arraybackslash\hspace{0pt}}m{#1}}
\newcolumntype{X}[1]{>{\raggedleft\let\newline\\\arraybackslash\hspace{0pt}}m{#1}}

\newcolumntype{R}[2]{%
    >{\adjustbox{angle=#1,lap=\width-(#2)}\bgroup}%
    l%
    <{\egroup}%
}

\providebool{fullversion}
\booltrue{fullversion}


\begin{document}

\title{Verifying Probabilistic Programs in Rust}

\author[A. Y. Bai]{Alexander Y. Bai}
\orcid{0009-0009-7458-7864}
\affiliation{
  \institution{New York University}
  \country{USA}
}
\email{ayb5065@nyu.edu}

\author[J. Tassarotti]{Joseph Tassarotti}
\orcid{0000-0001-5692-3347}
\affiliation{
  \institution{New York University}
  \country{USA}
}
\email{jt4767@nyu.edu}
\authornote{Also affiliated with Amazon Web Services. This paper does not reflect the views of Amazon Web Services.}

\begin{abstract}
Recent work has developed many techniques for formally verifying probabilistic programs.
However, existing verification frameworks for probabilistic programs are restricted to custom, idealized languages designed for verification.
As a result, they cannot be used to verify off-the-shelf probabilistic programs written in standard languages.
In contrast, for non-probabilistic programs, a number of verification tools now support verifying realistic code written in widely used languages such as Go, C, and Rust.
To verify probabilistic programs written in these languages, it would be useful to be able to reuse, as much as possible, the extensive development work that has gone into such tools.

This paper presents \thetool, a framework for verifying probabilistic Rust programs.
\thetool is based on Verus, a verification tool for Rust that supports SMT-based automation and separation-logic-inspired reasoning features.
\thetool extends Verus with support for probabilistic reasoning while retaining these expressive features.
 To do so, \thetool uses a lightweight encoding of probabilistic \emph{error credits}, a form of ghost state for randomized reasoning introduced in the Eris program logic.
By deriving an appropriate specification using error credits, \thetool supports verifying the correctness of randomized sampling algorithms.
We use this technique to verify several sampling routines for discrete distributions, including samplers for the discrete Laplace and discrete Gaussian distributions, the alias method, and the fast loaded dice roller.

We establish the soundness of our error credit extension by adapting VerusBelt, a recently developed logical relations model of Verus that encodes its features in terms of the Iris separation logic.
To do so, we replace the use of Iris's standard weakest precondition in this model with Eris's probabilistic weakest precondition instead.
The resulting soundness proof is fully mechanized in Rocq.
\end{abstract}

\maketitle

\section{Introduction}

Probabilistic programs are challenging to implement correctly, and their randomized behavior can make them difficult to test.
At the same time, bugs in probabilistic programs can have critical consequences, especially in  security applications, where randomness is an essential part of algorithms for cryptography and differential privacy.
Because of these important applications, there has long been interest in developing program logics to formally verify the correctness of probabilistic programs.
As a result, prior work has developed a wide range of probabilistic program logics~\citep{DBLP:series/mcs/McIverM05,ellora,psl,prhl_popl09, ub-icalp16, clutch}.
However, existing probabilistic program logics cannot be used to verify probabilistic programs written in standard, full-fledged programming languages, because at present, these logics only support special, idealized languages.
While these languages are suitable for modeling core aspects of important randomized programs and data structures, there is a gap between these models and real implementations.

Meanwhile, for \emph{non-probabilistic} programs, there is now a range of program-logic-based tools and frameworks for verifying complex programs written in languages such as Go, C, and Rust~\citep{verus-oopsla23, verus-sosp24, perennial, vst, gobra}.
These tools often come with sophisticated features that are needed to reason about the challenging patterns found in real-world programs, and have been used to verify substantial systems written in these languages.

In light of these successes, a natural question arises as to whether these non-probabilistic program logics can be extended to support probabilistic reasoning, or whether probabilistic program logics can be extended to support full-featured languages.
Unfortunately, both of these routes are challenging to carry out with many of the approaches that are commonly used to construct probabilistic program logics.
The core issue is that many probabilistic logics are structured in a way that is radically different from how standard non-probabilistic verification tools work.
For example, logics based on weakest preexpectation transformers~\citep{DBLP:series/mcs/McIverM05} change program logic assertions so that they are no longer predicates on program states, but are instead functions from program states to real numbers.
In another direction, other probabilistic program logics such as Ellora~\cite{ellora}, PSL~\cite{psl}, and Lilac~\cite{lilac} make assertions into predicates over \emph{distributions} of program states.
Reconciling these foundations with the approaches that non-probabilistic verification tools use for automation and support for reasoning about heaps and pointers is an open research problem.
Another issue is that many modern non-probabilistic program logics make use of rich forms of \emph{ghost state}.
Despite promising recent efforts to develop analogous theories of ghost state for probabilistic program logics~\citep{amaryllis}, the current state of the art for probabilistic ghost state still does not have the full flexibility found in non-probabilistic verification.

However, there is one class of program logics for probabilistic verification that appears to be easier to reconcile with state-of-the-art non-probabilistic program verification techniques.
These \emph{lifting-based} logics do \emph{not} change the type of assertions.
Instead, some limited aspect of a program's probabilistic behavior is tracked through a mechanism like ghost state.
Then, just the weakest precondition or Hoare triple of the logic is altered to give this ghost state a probabilistic interpretation. 
This approach was pioneered by pRHL~\citep{prhl_popl09}, where it was used for relational probabilistic reasoning.
Subsequent lines of work have applied the lifting-based approach in a number of program logics~\citep{ub-icalp16, clutch, approxis, tachis} for both unary and relational properties.
Some of these logics have recently been based on the Iris separation logic framework, showing that the lifting-based approach can combine the expressive separation logic features of Iris with probabilistic reasoning.
Still, that prior work has focused on a toy, idealized Mini-ML-like language, and cannot be used to reason about actual executable programs written in a realistic language.

This paper presents \thetool, a verification framework for reasoning about probabilistic programs written in Rust.
\thetool is based on Verus~\citep{verus-oopsla23, verus-sosp24}, a widely used SMT-based semi-automated verification tool for Rust. 
\thetool uses a lifting-based approach to incorporate probabilistic reasoning into Verus by extending Verus with \emph{probabilistic error credits}, a form of ghost state for tracking probabilities of events that was introduced in the prior Eris program logic~\citep{eris}.
With error credits, program specifications written in \thetool can be used to bound the probability that a program's execution fails to satisfy some property.
By proving a specification of an appropriate form, one can use error credits to show that an implementation of a routine for sampling from some probability distribution correctly generates samples with the right probabilities.

\thetool provides support for proving that a probabilistic program terminates \emph{almost surely}, \ie, terminates with probability 1.
This form of termination reasoning is known to be challenging.
While Verus provides built-in support for proving termination by annotating recursive functions and loops with some well-founded decreasing measure, many almost-surely terminating probabilistic programs have no such obvious decreasing measure.
Prior work on program logics and deductive verifiers for almost-sure termination has come up with alternate, subtle proof rules for working around this issue~\citep{McIverMKK18,heyvl}, but adapting Verus to support these alternate rules would require challenging engineering effort.
We are able to sidestep this issue entirely: \citet{eris} show an alternate method of establishing almost-sure termination that uses error credits themselves as the object to induct on, and \thetool is able to use an analogous technique with Verus's existing \texttt{decreases} clauses.

Because Verus supports rich forms of ghost state, the embedding of error credits in \thetool{} is relatively lightweight, and only requires adding two axioms to Verus.
The first is a specification for a library method for generating random integers, which connects the resulting random samples to the error credits.
The second is a direct translation of one of the primitive rules for error credits found in Eris.
On top of this, reasoning about probabilities of events uses Verus's recent support for Z3's real number theory, which we augment with a small library of results about discrete sums.

Although this axiomatic extension is relatively small, one might still wonder whether it is sound.
Therefore, to justify the soundness of \thetool's error credit encoding, we build on the recent VerusBelt project~\citep{verusbelt}, which constructs a semantic model of a large subset of Verus through an encoding into the Iris separation logic.
At a high level, the VerusBelt model extends the earlier RustBelt~\citep{rustbelt} model of Rust's type system by incorporating Verus's specification extensions to Rust's types.
The model establishes that if a Rust program (written in a core subset) is well-typed using Verus's extensions, then a corresponding Iris weakest precondition assertion holds.
Thus, the soundness of Verus (or at least the modeled subset) follows from the soundness of Iris.
We adapt this model to include \thetool's types for error credits and the proof rules they support.
To do so, we replace the use of ``standard'' Iris weakest precondition in the VerusBelt model with the Eris program logic's weakest preconditions instead.
This requires generalizing Eris in several ways, including extending it to cover the core Rust-like language used in VerusBelt, as well as incorporating certain Iris features that were missing from the original Eris.
Just as the original VerusBelt shows that whenever a program has a certain Verus specification, a corresponding Iris weakest precondition must hold, our adaptation shows that whenever a program satisfies a specification using \thetool's extensions, a corresponding Eris weakest precondition holds.
The soundness theorem of Eris then transfers to such programs.
This semantic model is fully mechanized in Rocq.

Finally, we demonstrate \thetool by using it to verify a number of sampling algorithms for discrete distributions written in Rust.
These include the OpenDP~\citep{opendp} differential privacy library's implementation of samplers for the discrete Laplace and discrete Gaussian, as well as the Fast Loaded Dice Roller~(FLDR)~\citep{fldr} and an efficient implementation of Walker's Alias Method~\citep{vose_alias, walker_alias, schwarz_blog}.
These first two examples demonstrate \thetool's ability to verify realistic samplers from security-critical applications, while the latter two involve stateful preprocessing, which works well with Verus's existing support for reasoning about state manipulation.
In addition to using Verus's Z3 automation, we employ Claude Opus 4.7 and 4.8, which are effective at constructing Verus proofs and can handle the mathematical reasoning about probabilities in these examples.

\paragraph{Contributions} To summarize, the contributions of this paper are:
\begin{itemize}[topsep=0pt]
\item \thetool, the first verification framework for probabilistic programs that supports verifying off-the-shelf programs written in a realistic, modern programming language.
\item A mechanized soundness proof for a core subset of \thetool, based on an adaptation of the VerusBelt semantic model.
\item A substantial set of examples of sampling algorithms verified using \thetool.
\end{itemize}

\section{Background}
\label{sec:background}

This section gives a brief overview of Verus and then explains how Eris's error credits work.

\subsection{Verus}
\label{sec:verus-bg}

Verus is a semi-automated verification tool for Rust programs~\cite{verus-oopsla23,verus-sosp24}.
Like Dafny~\cite{leino2010dafny}, it generates verification conditions that are discharged by an SMT solver.
The developer annotates functions with \texttt{requires}/\texttt{ensures} contracts and
loops with \texttt{invariant}s, and Verus solves the resulting proof obligations 
automatically with Z3~\cite{moura2008z3}. 
Code is partitioned into three modes: \texttt{exec} code that compiles and runs, 
\texttt{spec} code that provides pure mathematical definitions for use in specifications, 
and \texttt{proof} code for writing lemmas and helping the SMT solver find proofs.
The \texttt{spec}- and \texttt{proof}-mode code is erased before compilation,
so verification imposes no runtime cost.
For example, in the snippet below, the \texttt{spec} functions \texttt{divides} and \texttt{is\_prime}
give pure mathematical definitions;
the \texttt{proof} function \texttt{even\_gt\_2\_isnt\_prime} is a lemma Z3 discharges from
those definitions; 
and the executable \texttt{is\_prime\_impl} is checked by Verus to satisfy the
specification given by its preconditions and postconditions.

\begin{lstlisting}[language=Verus,style=VerusLineNos,columns=fixed,literate={}]
spec fn divides(n: int, k: nat) -> bool { n %

spec fn is_prime(n: nat) -> bool { forall|k: nat| 2 <= k < n ==> !divides(n as int, k) }

proof fn even_gt_2_isnt_prime(i: nat)
  requires i > 2 && is_even(i as int)
  ensures !is_prime(i) { assert(divides(i as int, 2)); }

fn is_prime_impl(n: u64) -> (result: bool) // only this exec fn will not be erased
  requires n >= 2,
  ensures result == is_prime(n as nat)
{ /* ... implementation and proof ... */ }
\end{lstlisting}
A distinguishing feature of Verus is how it reasons about \emph{unsafe} Rust. Safe Rust forbids
aliased mutable state, but systems code routinely escapes that fragment through raw pointers and
interior mutability, despite being semantically safe. 
Verus reasons about this unsafe code by pairing mutable state with a \emph{permission} token. 
For example, a permissioned pointer of type \texttt{PPtr<T>} cannot be written to directly on its own.
Instead, the right to access the underlying data 
is a separate ghost token \texttt{PointsTo<T>}, which carries the location's ghost value.
This is the Verus analogue of the separation-logic points-to assertion $\ell \mapsto v$.
Reading or writing through the pointer consumes and returns this token in the operation's \texttt{requires} and \texttt{ensures} clauses.
For instance, allocating a heap cell of type \texttt{PPtr<u64>} returns the raw pointer together with
its \texttt{PointsTo} token, and every access threads that token through the operation's contract:

\begin{lstlisting}[language=Verus,style=VerusLineNos,columns=fixed,literate={}]
let (ptr, Tracked(mut perm)) = PPtr::<u64>::new(7); // ptr |-> 7;  perm is the token
let x = ptr.read(Tracked(&perm));                   // read borrows the token:  x == 7
ptr.write(Tracked(&mut perm), x + 1);               // write mutates it: now ptr |-> 8
let z = ptr.take(Tracked(&mut perm));               // move the value out: ptr |-> uninit; z == 8
ptr.free(Tracked(perm));                            // free consumes the token
\end{lstlisting}

The \texttt{PointsTo} token is just one instance of a more general facility for tracking logical permissions associated with state.
Specifically, variables in Verus have three modes: \texttt{ghost}, \texttt{tracked}, and \texttt{exec}. The first two
are erased at compile time and are used for specifications and proofs. The \texttt{ghost} values are
duplicable, while \texttt{tracked} values, like \texttt{PointsTo}, are \emph{affine}.
This means they cannot be copied, and the type checker tracks the transfer of their ownership through the program just
like standard Rust ownership of physical state. A function consumes the \texttt{tracked} arguments it is passed and must
return any it intends to give back. This makes \texttt{tracked} state behave
like a separation-logic resource.
While Verus uses Rust's substructural type system to manage these affine resources, the SMT solver checks the validity of specifications \cite{hance-thesis,verusbelt}.

 As in modern separation logic frameworks like Iris, Verus allows developers to define custom forms of ghost state that can be used to encode appropriate forms of permissions~\citep{hance-thesis, verusbelt}.
This ghost state needs to support various operations, which are required to satisfy certain algebraic laws, like the \emph{resource algebras} used for ghost state in Iris.
Concretely, Verus provides a two-layer interface for defining new ghost state:
a \texttt{ResourceAlgebra} trait with a composition \texttt{op} and a validity
predicate \texttt{valid}, along with the associated algebraic laws, and a \texttt{PCM} trait that additionally provides a
\texttt{unit}. To create a new form of ghost state, the user chooses a carrier type \texttt{P},
and implements these traits.

\begin{lstlisting}[language=Verus,style=VerusLineNos,columns=fixed,literate={}]
trait ResourceAlgebra: Sized { 
    spec fn valid(self) -> bool;
    spec fn op(a: Self, b: Self) -> Self;

    proof fn associative(a: Self, b: Self, c: Self)
        ensures Self::op(a, Self::op(b, c)) == Self::op(Self::op(a, b), c);
    proof fn commutative(a: Self, b: Self)
        ensures Self::op(a, b) == Self::op(b, a);
    proof fn valid_op(a: Self, b: Self)
        requires Self::op(a, b).valid(), ensures a.valid();
}

trait PCM: ResourceAlgebra {
    spec fn unit() -> Self;                    // identity for op
    proof fn op_unit(self) ensures Self::op(self, Self::unit()) == self;
    proof fn unit_valid() ensures Self::unit().valid();
}
\end{lstlisting}

After establishing the \texttt{PCM} trait for a type \texttt{P}, the user can then instantiate the \texttt{tracked} type \texttt{Resource<P>}.
These resources live at a ghost location \texttt{loc()}, and come equipped with proof functions \texttt{alloc} (creates a new piece of ghost state), \texttt{join}/\texttt{split} (composition and decomposition via \texttt{op}), and \texttt{validate} (which establishes that the owned value satisfies \texttt{valid}).
This ghost state can be modified using \texttt{update}, which allows the ghost state to be changed in a \emph{frame-preserving} way, meaning that the updated ghost state must be compatible with other possible parts of that ghost state that might be owned elsewhere.
Having such a \texttt{Resource<P>} in a context is analogous to owning the corresponding separation-logic resource assertion.

Another important feature of Verus is its support for reasoning about termination. This feature 
is necessary for \texttt{spec} and \texttt{proof} code to prevent unsoundness from 
circular reasoning~\cite{verus-oopsla23}. Verus checks termination by requiring that 
recursive functions and loops are annotated with a \texttt{decreases} clause, 
which specifies a well-founded measure that decreases on each recursive call or loop iteration.
For example, the following recursive \texttt{spec} function \texttt{factorial} is accepted
because its argument \texttt{n} is a natural number that strictly decreases on each recursive call,
making it a valid well-founded measure:
\begin{lstlisting}[language=Verus,style=VerusLineNos]
spec fn factorial(n: nat) -> nat
    decreases n,
{
    if n == 0 { 1 } else { n * factorial((n - 1) as nat) }
}
\end{lstlisting}

Finally, Verus offers flexible ways to attach trusted specifications to executable code.
This flexibility is useful for modeling external aspects of the execution environment or library functions.
A function marked \texttt{\#[verifier::external\_body]}
has a compiled body that Verus treats as opaque (thus ``external'' to Verus),
and its specification is trusted. For example, this is how raw-pointer access itself is given a specification.
The \texttt{ptr\_ref} function dereferences a \texttt{*const T}, and its trusted contract uses the \texttt{PointsTo}
permission to justify the \texttt{unsafe} body.
\begin{lstlisting}[language=Verus,style=VerusLineNos,columns=fixed,literate={}]
#[verifier::external_body]
pub fn ptr_ref<T>(ptr: *const T, Tracked(perm): Tracked<&PointsTo<T>>) -> (v: &T)
    requires
        perm.ptr() == ptr, perm.is_init(),   // must point to initialized memory
    ensures
        v == perm.value(),                   // the borrow reads that value
    opens_invariants none
    no_unwind
{
    unsafe { &*ptr }                         // real unsafe Rust, unchecked by Verus
}
\end{lstlisting}
The specification we write for \texttt{\#[verifier::external\_body]} extends the trusted computing base
(TCB), and the developer takes on the
obligation that the implementation actually meets it. 
Verus includes axiomatized rules for primitives like \texttt{PPtr} and \texttt{PCell} in this way.
The VerusBelt project~\citep{verusbelt} subsequently constructed a soundness proof to justify these axioms.
As we will see, \thetool{} uses the same facility to axiomatize aspects of the probabilistic ghost state it uses.

\subsection{Eris}
\label{sec:eris}

As mentioned in the introduction, Eris is a separation logic for proving upper bounds on error probabilities.
The language targeted by Eris is an idealized, probabilistic Mini-ML-like language called \textsc{ProbLang}.
Roughly speaking, one can think of this language as a sequential version of the default \textsc{HeapLang} that ships with Iris, extended with a command for generating random samples.
\citet{eris} present two different versions of the Eris program logic: one for partial correctness and one for total correctness.
For compatibility with Verus's support for proving termination, we will focus on the total-correctness version.

The key idea behind Eris is the introduction of \emph{probabilistic error credits}, a separation logic resource that can be used to track upper bounds on error probabilities.
Tracking such probabilities is useful because many probabilistic programs and algorithms have some probability of failure.
For example, the Miller-Rabin primality test~\citep{rabin, miller} has some probability of incorrectly declaring that a composite number is prime.
Other examples show up in cryptography and differential privacy, where a security claim holds except when a rare event occurs, such as a hash collision or an attacker guessing a randomly generated key.

When reasoning about such algorithms, a key goal is to prove an upper bound on this probability of failure.
Eris's error credit assertions, which have the form $\upto{\err}$ for $0 \leq \err \leq 1$, represent a logical permission to execute actions that might cause an error with probability at most $\err$. 
Using the logic, one proves total Hoare triple specifications of the form $\thoare{\upto{\err} \sep \prop}{e}{v.\; Q(v)}$.
A triple of this form says that if we execute $e$ in a state initially satisfying $P$, then with probability at least $1 - \err$, $e$ will terminate with a value $v$ satisfying $Q(v)$.
In other words, the initial ``budget'' of $\upto{\err}$ error credits in the precondition licenses $e$ to ``fail'' with probability at most $\err$, where a failing execution involves either (1) violating safety and getting ``stuck''; (2) returning a value $v$ for which $Q(v)$ does not hold; or (3) not terminating.

Error credits are used in the proof rules of the logic to exclude reasoning about certain cases or branches of execution in which an error would occur.
For example, the following proof rule reasons about a $\texttt{rand}(N)$ command, which samples an integer uniformly from the set $\{0, \dots, N - 1\}$:%
\footnote{In \citet{eris}, this command samples from the set $\{0, \dots, N\}$ instead of $\{0, \dots, N-1\}$. We adopt the alternative convention here to match the form we later use in Rust.}
\begin{mathpar}
  \infrule[lab]{Err-Rand-Spend}
    { |S| = M }
    {\proves \thoare{\upto{\frac{M}{N}}}{\texttt{rand}(N)}{v.\; v \notin S}}
\end{mathpar}
This rule allows the user to pick a set of integers $S$ of size $M$, and in the postcondition we are guaranteed that $v$ is \emph{not} in that set $S$.
That is, we have excluded reasoning about the cases where the returned integer is in the set $S$.
Doing so requires $M/N$ error credits from the precondition, because the probability of drawing an element in $S$ is at most $M/N$, since each number $i$ in $\{0, \dots, N-1\}$ is selected with probability $1/N$.
Because Eris is a separation logic, applying the rule consumes the error credits from the precondition.

In Eris, this rule is in fact a derived rule.
At its core, the logic provides just the following four primitive proof rules for error credits:

\begin{mathpar}
  \infrule[lab]{Err-Split}{\err_1 \geq 0 \\ \err_2 \geq 0}{\upto{\err_1 + \err_2} \dashv\vdash \upto{\err_1} \ast \upto{\err_2}}
    
  \infrule[lab]{Err-1}{}{\upto{1} \vdash \FALSE}

  \infrule[lab]{Err-Thin-Air}
    {\forall \err > 0.\; \proves \thoare{\upto{\err} \sep \prop}{e}{v.\; \Phi(v)}}
    {\proves \thoare{\prop}{e}{v.\; \Phi(v)}}
    
  \infrule[lab]{Err-Rand-Exp}
    { \forall i < N.\; 0 \le \Err(i) \le 1 \\
      \tfrac{1}{N}\sum_{i=0}^{N-1} \Err(i) = \err
    }
    {\proves \thoare{\upto{\err}}{\texttt{rand}(N)}{v.\; \upto{\Err(v)}}}
\end{mathpar}

The \ruleref{Err-Split} rule allows for splitting and joining error credits across the separating conjunction.
This is useful because it allows us to divide up a budget of error credits and pass ownership of a part of the budget to different components or modules that need them.
\ruleref{Err-1} allows us to derive $\FALSE$ once we have an error credit of $1$.
Intuitively, this follows because, if we read $\upto{\err}$ as permission to fail to satisfy a specification with probability $\err$, then when $\err = 1$, we can fail to satisfy the specification with probability $1$, so there is nothing to prove.
Reading \ruleref{Err-Thin-Air} from the bottom up, the rule allows us to extend the precondition by some additional error credit $\err > 0$.
The quantification here says that this extra credit we produce out of ``thin air'' may be an arbitrarily small positive number.
The soundness of this rule follows from a kind of continuity property of the Hoare triple, where we take the limit as $\err \rightarrow 0$.

Finally, \ruleref{Err-Rand-Exp} is what connects error credits to random sampling with $\texttt{rand}(N)$. 
When applying this rule, the user picks a function $\Err : \{0, \dots, N-1\} \to [0, 1]$, which maps outcomes of the random sampling to numbers in $[0,1]$.
It then says that if we start with an initial budget of $\err$ credits, then when $\texttt{rand}(N)$ returns $v$, we will end up with $\Err(v)$ error credits in the postcondition.
The premise requires that the \emph{expected value} of $\Err$ across the random outcomes is equal to the initial budget $\err$ of error credits we started with.
Since $\texttt{rand}(N)$ returns an integer uniformly from the set $\{0, \dots, N-1\}$, this expected value is computed by taking the average of $\Err$ across the outcomes.
In particular, by using rule \ruleref{Err-Rand-Exp}, we can derive the earlier \ruleref{Err-Rand-Spend} by distributing the credits so that we have one credit on the branches where $v \in S$ and zero credits on the other branches. Then, by applying rule \ruleref{Err-1}, we can spend the one error credit on the branches where $v \in S$ to derive $\FALSE$, and thus exclude any further obligations on those branches.

\paragraph{Sampler Correctness}
\label{par:sampler-correctness}

At first, it might appear that Eris only allows for proving relatively limited kinds of specifications, because many properties of interest cannot be expressed merely as upper bounds on error probabilities.
However, it turns out that appropriately bounding error probabilities suffices to completely characterize the distribution of values that a program can return.
In particular, \citet{cpp26-distribution} proved a stronger soundness theorem for Eris that allows one to prove that a program draws samples from a given distribution.
Their key idea is that \ruleref{Err-Rand-Exp} encodes the fact that the $\texttt{rand}(N)$ command samples uniformly from $\{0, \dots, N - 1\}$ by controlling how credits may be redistributed.
Thus, by proving a similar credit redistribution specification about a program $e$, we can specify what distribution of values $e$ generates.
Specifically, to prove that a program $e$ samples from distribution $\distr$, their soundness theorem says that it suffices to prove a specification of the form:
\[
  \infrule<EPT>[lab]{Expectation Preserving Transformation (EPT)}
  { %
    \All v \in \Val\ .\ 0 \le \Err_2(v) \le 1 \\
    \sum_{v \in \Val} \distr(v) \cdot \Err_2(v) = \err_1
  }
  {\proves \thoare{\upto{\err_1}}{e}{v\ldotp\ \upto{\Err_2(v)}}}
\]
where the sum in the premise is computing the expected value of $\Err_2$ under distribution $\distr$, and the rule requires this expected value to be equal to $\err_1$.

\section{Adding Eris-Style Reasoning to Verus}
\label{sec:verus}

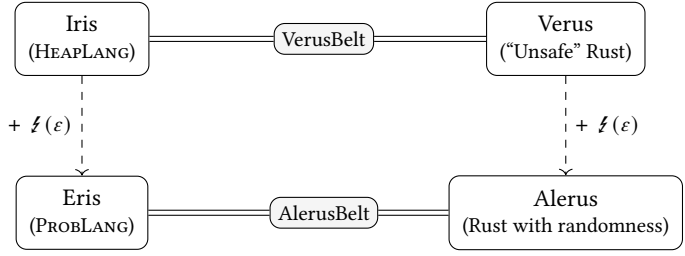
\begin{figure}[h]
\centering
\begin{tikzpicture}[
  box/.style={draw, rounded corners, align=center, inner sep=5pt, font=\small},
  bridge/.style={draw, rounded corners, fill=black!4, inner sep=3pt, font=\footnotesize},
  conn/.style={double, double distance=1.5pt},
  lbl/.style={font=\footnotesize},
]
  \node[box] (iris)  at (0,0)    {Iris\\[-1pt]{\footnotesize (\textsc{HeapLang})}};
  \node[box] (verus) at (6.4,0)  {Verus\\[-1pt]{\footnotesize (``Unsafe'' Rust)}};
  \node[box] (eris)  at (0,-2.3) {Eris\\[-1pt]{\footnotesize (\textsc{ProbLang})}};
  \node[box] (veris) at (6.4,-2.3) {\thetool\\[-1pt]{\footnotesize (Rust with randomness)}};

  \node[bridge] (vb)  at (3.2,0)    {VerusBelt};
  \node[bridge] (ab)  at (3.2,-2.3) {AlerusBelt};
  \draw[conn] (iris.east) -- (vb.west);
  \draw[conn] (vb.east)   -- (verus.west);
  \draw[conn] (eris.east) -- (ab.west);
  \draw[conn] (ab.east)   -- (veris.west);

  \draw[->,dashed] (iris)  -- node[lbl,left]{$+\;\upto{\err}$} (eris);
  \draw[->,dashed] (verus) -- node[lbl,right]{$+\;\upto{\err}$} (veris);
\end{tikzpicture}
\caption{High-Level Picture of \thetool}
\label{fig:analogy}
\end{figure}

\thetool\ adds Eris's error credits to Verus without modifying the verifier.
Just as Verus uses \texttt{tracked} ghost state to implement permissions analogous to separation logic's points-to, \thetool{} uses tracked ghost state to represent Eris's error credits.
The Eris reasoning rules are then encoded as a small library of proof functions over these resources.
To justify the soundness of this encoding, we construct a semantic model that relates it to Eris, just as VerusBelt relates Verus's encoding of permissions to Iris.
This analogy between the designs is illustrated in \Cref{fig:analogy}.
This section describes the encoding. The semantic model is later explained in \Cref{sec:verisbelt}.

\subsection{Error Credits in Verus}

We obtain error credits by implementing the \texttt{ResourceAlgebra} and \texttt{PCM} interfaces of
\Cref{sec:background} for the following carrier type:
\[
  \mathsf{ErrorCreditCarrier} \;::=\; \mathsf{Value}(c \in \mathbb{R}) \;\mid\; \mathsf{Empty} \;\mid\; \mathsf{Invalid}
\]
A $\mathsf{Value}(c)$ represents an error credit of magnitude $c$.
Note that the $\mathsf{Value}$ type ranges over arbitrary real numbers instead of merely non-negative real numbers.
While negative error credits should not be representable, we use arbitrary reals so that we can interface with the SMT real theory, which works over arbitrary reals.
Meanwhile, $\mathsf{Empty}$ is the \texttt{PCM} unit, analogous to having zero error credits.
Finally, as the name suggests, $\mathsf{Invalid}$ is used to represent combinations of error credit resources that are unrepresentable or invalid.

The composition \texttt{op}, written as $\cdot$, is defined by
\[
  \mathsf{Value}(c_1) \cdot \mathsf{Value}(c_2) =
  \begin{cases}
    \mathsf{Value}(c_1 + c_2) & \text{if } c_1 \ge 0 \text{ and } c_2 \ge 0, \\
    \mathsf{Invalid}          & \text{otherwise.}
  \end{cases}
\]
A credit is valid when it is either an empty credit or a value of $c$ for $c$ lying in the range $[0, 1)$:
\[
  \mathsf{valid}(\mathsf{Value}(c)) \iff 0 \le c < 1.
\]
All error credits live at a single global ghost location.
A \texttt{tracked ErrorCreditResource} whose view is $\mathsf{Value}(\err)$ is then analogous to
the Eris assertion $\upto{\err}$.

Then, Eris's \ruleref{Err-Split}, which allows for credit composition $\upto{\err_1} * \upto{\err_2} \;\dashv\vdash\; \upto{\err_1+\err_2}$,
is witnessed by the proof functions \texttt{ec\_combine} and \texttt{ec\_split}, which respectively merge two
credits and split one credit into two summands.
These have the following signatures:
\begin{center}
\begin{minipage}[t]{0.49\linewidth}
\begin{lstlisting}[language=Verus,style=VerusLineNos]
pub proof fn ec_combine(
    tracked c1: ErrorCreditResource,
    tracked c2: ErrorCreditResource,
    v1: real, v2: real,
) -> (tracked out: ErrorCreditResource)
    requires
        c1@ =~= Value { car: v1 },
        c2@ =~= Value { car: v2 },
        v1 >= 0real, v2 >= 0real,
    ensures
        out@ =~= Value { car: v1 + v2 },
\end{lstlisting}
\end{minipage}\hfill
\begin{minipage}[t]{0.49\linewidth}
\begin{lstlisting}[language=Verus,style=VerusLineNos]
pub proof fn ec_split(
    tracked c: ErrorCreditResource,
    v1: real, v2: real,
) -> (tracked (c1, c2): 
    (ErrorCreditResource, ErrorCreditResource))
    requires
        c@ =~= Value { car: v1 + v2 },
        v1 >= 0real, v2 >= 0real,
    ensures
        c1@ =~= Value { car: v1 },
        c2@ =~= Value { car: v2 },
\end{lstlisting}
\end{minipage}
\end{center}

Both are pure \texttt{proof} functions, so they manipulate only ghost state and are erased
after verification.
The \texttt{ec\_combine} function consumes two \texttt{tracked} resources whose views, 
represented by Verus's postfix \texttt{@} operator, 
are $\mathsf{Value}(v_1)$ and $\mathsf{Value}(v_2)$ and returns a single resource 
whose view is $\mathsf{Value}(v_1 + v_2)$, realizing the right-to-left direction of \ruleref{Err-Split}.
Meanwhile, \texttt{ec\_split} runs this in reverse: it consumes one resource of view $\mathsf{Value}(v_1 + v_2)$ and hands back a pair of resources with views $\mathsf{Value}(v_1)$ and $\mathsf{Value}(v_2)$.
These are \texttt{Resource}'s \texttt{join} and \texttt{split} functions for this instance.
In both directions the $v_i \ge 0$ preconditions mirror the non-negativity side condition built into \texttt{op}, so that the credits being merged or divided are always valid; the $\mathtt{=\!\sim\!=}$ relation is Verus's extensional equality on views.

\begin{wrapfigure}{r}{0.34\textwidth}
\begin{lstlisting}[language=Verus,style=VerusLineNos]
pub proof fn ec_contradict(
    tracked e: &ErrorCreditResource,
)
    requires
        exists |car: real| car >= 1real
            && e@ =~= Value { car },
    ensures
        false,
\end{lstlisting}
\label{fig:ec-contradict}
\end{wrapfigure}
Next, we get something analogous to \ruleref{Err-1}, which says $\upto{1} \vdash \bot$, through the proof function \texttt{ec\_contradict}.
This derives a contradiction from any credit of value greater than or equal to one, using
\texttt{Resource}'s \texttt{validate} to learn that the held value is \texttt{valid}.
Here, \texttt{ec\_contradict} takes only a shared reference to a credit resource, since it does not consume the credit but merely inspects it.
Its precondition asserts that the resource's view is some $\mathsf{Value}(\mathit{car})$ with $\mathit{car} \ge 1$, and from this it derives $\mathsf{false}$.
The proof of this follows from the fact that Verus already has a rule saying that ownership of an invalid element implies false.
Since a credit is valid only when its value lies in $[0, 1)$, ownership of a credit of value $\ge 1$ thus implies false.

\subsection{Axiomatized Eris Rules in Verus}
\label{sec:axioms}

The algebraic rules for error credits shown above are proved from the resource definition.
The two remaining Eris rules, expectation preservation (\rref{Err-Rand-Exp}) and thin air (\rref{Err-Thin-Air}),
are the only constructs \thetool\ adds as trusted axioms. 
Both are axiomatized as executable functions acting on the credit resource.
This is similar to how Verus must axiomatize the connection between points-to resources and the underlying Rust commands that mutate physical state, such as \texttt{ptr\_mut\_ref}.

\paragraph{Expectation Preservation.} 
To encode the expectation-preserving rule, we first need to decide on the primitive that we want to use for drawing random samples.
In our examples, we use a single primitive for drawing random numbers,
\texttt{rand\_ubig}, which generates random samples and returns values of type \texttt{UBig},
which represents an arbitrary-precision unsigned integer.
We also use \texttt{rand\_ubig} to derive \texttt{rand\_u64}, which generates uniform random samples over 64-bit unsigned integers.\footnote{For performance reasons, you might want to have \texttt{rand\_u64} as a standalone axiomatized primitive, but we do not do so here for simplicity.}
The bignum implementation is from the \texttt{dashu} crate~\cite{dashu}.
We give the bignum operations trusted specifications, since verifying the bignum library itself
is beyond the scope of this paper.
\begin{lstlisting}[language=Verus,style=VerusLineNos]
#[verifier::external_body]
pub fn rand_ubig(
    bound: &UBig,
    Tracked(e1): Tracked<ErrorCreditResource>,
    Ghost(e2): Ghost<spec_fn(nat) -> real>,
) -> ((n, out_credit): (UBig, Tracked<ErrorCreditResource>))
    requires
        ubig_view(bound) > 0,
        forall |i: nat| e2(i) >= 0real,
        exists |eps: real| e1@ =~= Value { car: eps }
            && eps >= average_nat(ubig_view(bound), e2),
    ensures
        ubig_view(&n) < ubig_view(bound),
        out_credit@@ =~= Value { car: e2(ubig_view(&n)) },
\end{lstlisting}

The function \texttt{rand\_ubig(bound, e1, e2)} samples a value $n$ uniformly from $\{0,\dots,N-1\}$, where
$N = \texttt{ubig\_view(bound)}$ is the arbitrary-precision bound. The \texttt{tracked} argument
\texttt{e1} carries the input credit of value $\err$, and the \texttt{Ghost} argument \texttt{e2} is the
credit-allocation function $\Err : \mathbb{N} \to \mathbb{R}$. The precondition requires
$\texttt{e2}(i) \ge 0$ for every $i$, together with the expectation-preservation side condition that the
held credit dominate the uniform average of \texttt{e2}, computed by \texttt{average\_nat}:
$\err \ge \texttt{average\_nat}(N, \texttt{e2}) = \tfrac1N\sum_{i<N}\Err(i)$. Just as in
\ruleref{Err-Rand-Exp}, the function returns the sample \texttt{n} together with the output credit
\texttt{out\_credit}, and the postcondition guarantees that this credit has value $\Err(n)$.

\paragraph{Thin Air.} The \texttt{thin\_air()} axiom returns a \texttt{tracked} credit of \emph{some} value $\varepsilon > 0$ out of thin air. 
Note that our encoding of this axiom deviates slightly from the way the rule \rref{Err-Thin-Air} was phrased: rather than quantifying over all possible $\varepsilon > 0$ and requiring that the Hoare triple be proved for each $\varepsilon$, we instead existentially quantify over some $\varepsilon > 0$ and add that to the context.
This formulation is logically equivalent, and is a better fit because Verus doesn't have impredicative Hoare triples, so there's no way for us to write a rule that requires a collection of Hoare triples to be proved.
\begin{lstlisting}[language=Verus,style=VerusLineNos]
#[verifier::external_body]
pub fn thin_air() -> (ret: Tracked<ErrorCreditResource>)
    ensures
        exists |eps: real| eps > 0
            && ret@@ =~= Value { car: eps },         // owns a credit eps > 0
\end{lstlisting}
One other difference between the formulation of this rule in Verus and the version in Eris is that this version requires that the program be explicitly annotated with the invocation of this ghost operation. In contrast, in Eris, the proof rule can be invoked at any point, without annotating the program.
In our examples, this difference is not a limitation, because there is usually a clear point where one wishes to invoke the rule, making it easy to add this annotation.

\subsection{Example}
\label{sec:verus-example}

We now put these pieces together in a small self-contained example.
Consider a program that flips two fair coins using
\lstinline[language=Verus]|rand_2_u64|,
and returns whether both came up
heads, where heads is encoded as $1$.
We will prove that, given $1/4$ error credit up front, the function
returns \texttt{false}.
In other words, the outcome where it would return \texttt{true} occurs with probability at most $1/4$.

\begin{lstlisting}[language=Verus,style=VerusLineNos]
pub fn flip_and(Tracked(credit): Tracked<ErrorCreditResource>) -> (ret: bool)
    requires
        credit@ =~= (Value { car: 1real / 4real }),
    ensures
        ret == false,
{
    let (b1, Tracked(c1)) = rand_2_u64(
        Tracked(credit), 
        Ghost(|x: nat| if x == 1 { 1real / 2real } else { 0real })
    );
    let (b2, Tracked(c2)) = rand_2_u64(
        Tracked(c1), 
        Ghost(|x: nat| if b1 == 1 && x == 1 { 1real } else { 0real })
    );
    proof { if b1 == 1 && b2 == 1 { ec_contradict(&c2); } }
    (b1 == 1) && (b2 == 1)
}
\end{lstlisting}

The proof works by picking the error credit allocation ($\Err$ in \ruleref{Err-Rand-Exp}) for 
each coin flip so that we get $\upto{1}$ for the ``bad'' outcome we need to rule out (both coin flips yielding 1).

For the first flip we hand \rref{Err-Rand-Exp} the allocation $\Err_1$ which gives $\tfrac12$ on
outcome $1$ and $0$ on outcome $0$. The average of this allocation is $\tfrac14$, 
which is exactly the credit we start with.
By the postcondition of \texttt{rand\_2\_u64} we then own $\upto{\Err_1(\mathtt{b1})}$.
For the second flip the allocation $\Err_2$ is conditioned on the first coin, paying $1$ when
$\mathtt{b1} = 1$ and the new draw is $1$, and $0$ in all other cases. 
The average of this allocation is $\tfrac12$ when $\mathtt{b1} = 1$ and $0$ when $\mathtt{b1} = 0$, 
which is exactly the credit we own after the first flip. Hence, along the single path
$\mathtt{b1} = \mathtt{b2} = 1$ we end up owning
$\upto{\Err_2(1)} = \upto{1}$, and on every other path only $\upto{0}$.
The \texttt{proof} block invokes \texttt{ec\_contradict} (\ruleref{Err-1}) exactly on the former
path, turning the full credit into a proof of \texttt{false}, thereby discharging that proof branch.
Since the return value 
$(\mathtt{b1} = 1) \wedge (\mathtt{b2} = 1)$ is \texttt{true} only on that
excluded path, every other case returns \texttt{false}, discharging the postcondition.

\section{Almost Sure Termination with Error Credits}
\label{sec:ast}

As discussed in \Cref{sec:background}, Verus enables showing termination via \texttt{decreases} clauses. 
In probabilistic programs, rather than focusing on the usual characterization of termination, it is common to consider \emph{almost sure termination}, \ie, to show that the program terminates with probability 1.
To see why this is challenging, consider the following program, which generates samples from the geometric distribution. 
It repeatedly generates samples from $\{0, 1\}$ uniformly and counts the number of $1$s generated before the first $0$.
The repetition is done by recursively calling \texttt{geometric} in the case where a $1$ is generated.

\begin{wrapfigure}{r}{0.45\textwidth}
\begin{lstlisting}[language=Verus,style=VerusLineNos,]
pub fn geometric() -> (ret: UBig) {
    let val = rand_2_u64();  // fair coin
    if val == 0 {
        UBig::ZERO           // terminate
    } else {
        geometric() + UBig::ONE  // recurse
    }
}
\end{lstlisting}
\label{fig:geometric}
\end{wrapfigure}
This program terminates with probability 1, because the diverging execution that repeatedly samples 1 forever occurs with probability 0.
However, we cannot establish termination using a traditional \texttt{decreases} clause because there's no obviously decreasing measure to supply.
The function takes no arguments, and the recursive call in the
\texttt{else} branch is made directly, with no quantity that visibly shrinks
between calls.

To deal with this, previous probabilistic program logics and deductive verifiers based on them~\cite{heyvl, McIverMKK18}
encode specialized reasoning rules for almost-sure termination.
However, in \thetool, we achieve almost-sure termination reasoning without needing to add any specialized rules. The key is that once we have error credits, we can formulate a decreasing measure that will work with Verus's existing \texttt{decreases} clause.

\citet{eris} call this form of termination reasoning ``credit amplification''.
The idea behind credit amplification is to show that on non-terminating branches, when the program recurses or loops, the amount of credit owned increases and will eventually reach 1.
Since $\upto{1} \vdash \bot$, once the proof accumulates a credit of value 1, we are done.
Under the hood, by accumulating error credits in this fashion, we have effectively shown that the probability of non-termination goes to zero.

The general methodology is to find a way to bound the number of steps or repetitions that it will take to reach $\upto{1}$, and then use that number of steps as the measure to supply to the \texttt{decreases} clause.
Because the number of steps is some finite number that will decrease, this works with Verus's existing mechanisms for reasoning about termination through the \texttt{decreases} clause.

\begin{wrapfigure}{r}{0.5\textwidth}
\begin{lstlisting}[language=Verus,style=VerusLineNos]
pub fn geometric() -> (ret: UBig) {
    let Tracked(e) = thin_air();
    let ghost depth: nat;
    let ghost eps: real;
    proof {
        eps = choose |v: real| e@.value() == Some(v);
        archimedean_exp_growth(eps, 2real);
        depth = choose |k: nat| 
                    eps * pow(2real, k) >= 1real;
    }
    bounded_geometric(Tracked(e), Ghost(depth))
}
\end{lstlisting}
\end{wrapfigure}
We showcase this technique first by proving almost sure termination of the geometric sampler. 
We reformulate the program below in order to thread the error-credit resource through it.
The entry point
\lstinline[language=Verus]|geometric| owns no error credit yet, so it invokes
\lstinline[language=Verus]|thin_air()| to materialize a credit of arbitrary value $\err > 0$ as the ghost resource
\lstinline[language=Verus]|e|.
On each iteration, we will redistribute the credits so that the terminating branch receives 0 credits, and the non-terminating branch obtains twice the amount of credit.
Thus, if we start with $\err$ error credits, then after $n$ iterations, we will have $2^n \cdot \err$ error credits.
The proof uses the Archimedean property of the reals
to pick an $n$ such that $\err \cdot 2^n \ge 1$,
and hands both $\err$ and $n$ to a recursive helper function \lstinline[language=Verus]|bounded_geometric|.

The helper \lstinline[language=Verus]|bounded_geometric| does the actual sampling. The invariant
$\err \cdot 2^{\mathtt{depth}} \ge 1$ is carried in the precondition, and termination is justified
by \lstinline[language=Verus]|decreases depth|; when $\mathtt{depth} = 0$ the invariant forces $\err \ge 1$.
Each iteration calls \lstinline[language=Verus]|rand_2_u64| with a credit allocation that assigns $0$ on the
terminating branch and $2\err$ on the recursing branch, which satisfies \rref{Err-Rand-Exp} as
$(0 + 2\err)/2 = \err$. So $\err$ is amplified to $2\err$ and forwarded to the
recursive branch. There, $2\err \cdot 2^{\mathtt{depth}-1} = \err \cdot 2^{\mathtt{depth}} \ge 1$,
restoring the invariant at the smaller depth and closing the recursion.
\begin{lstlisting}[language=Verus,style=VerusLineNos]
pub fn bounded_geometric(Tracked(in_credit): Tracked<ErrorCreditResource>, Ghost(depth): Ghost<nat>) -> UBig
    requires exists |eps: real| {
        &&& eps > 0real
        &&& in_credit@ =~= (Value { car: eps })
        &&& eps * pow(2real, depth) >= 1real
    },
    decreases depth,
{
    /* ... get epsilon and show depth is non-zero ... */
    let (val, Tracked(out_credit)) = rand_2_u64(Tracked(in_credit),
        Ghost(|x: nat| if x == 0 { 0real } else { 2real * eps }), // 0 |-> 0,  1 |-> 2*eps
    );
    if val == 0 {
        UBig::ZERO  // terminating branch
    } else {
        // 2*eps * 2^(depth-1) = eps * 2^depth >= 1, invariant restored
        bounded_geometric(Tracked(out_credit), Ghost((depth - 1) as nat)) + UBig::ONE
    }
}
\end{lstlisting}

\paragraph{Beyond geometric loops.}
The almost-sure termination reasoning for the geometric sampler is quite simple: 
the credit consistently doubles in the non-terminating branch. 
To show we can also handle nontrivial almost-sure termination properties,
we next prove the almost-sure termination of a one-dimensional random walk.
This example will illustrate a general
methodology for using the credit allocation to make the termination proof go through.

The one-dimensional random walk is a probabilistic program that starts at any position
$n \in \mathbb{N}$; at each step, it increases the position by $1$ or decreases it by $1$ with equal probability.
The program terminates when it first hits $0$. 
Unlike in the case of the geometric sampler, here we cannot pick a constant amplification
factor for the error credits, since the walker can drift arbitrarily far from the origin before returning.

\begin{wrapfigure}{r}{0.45\textwidth}
\begin{lstlisting}[language=Verus,style=VerusLineNos]
pub fn random_walk(pos: UBig) -> (ret: UBig) {
    if pos == UBig::ZERO {   // at the origin
        UBig::ZERO
    } else {
        let val = rand_2_u64();  // fair coin
        if val == 0 {
            random_walk(pos - UBig::ONE)
        } else {
            random_walk(pos + UBig::ONE)
        }
    }
}
\end{lstlisting}
\label{fig:random-walk}
\end{wrapfigure}
The key is that the credit allocation function we hand to \texttt{rand\_2\_u64},
rather than amplifying the credit by a fixed amount, now depends on
the current position.
How do we define what this function should be?
Recall that we ultimately want to have some ``fuel'' parameter $s$ that decreases on each recursive call or loop iteration such that when $s = 0$, the program terminates.
For a given value of $s$, we can think of the credit allocation function as needing to distribute credits so that when the walker ends up in state $p$, we receive
credits equal to the probability that the program \emph{does not} terminate after $s$ rounds starting from $p$.

We can define this function $\mathtt{fail\_prob}(s, p)$ in terms of a recurrence relation:
\begin{equation}\label{eq:fail-prob}
\begin{aligned}
  \mathtt{fail\_prob}(s, 0) &\;=\; 0, \\
  \mathtt{fail\_prob}(0, p) &\;=\; 1 \quad \text{for } p > 0, \\
  \mathtt{fail\_prob}(s, p) &\;=\; \tfrac{1}{2}\bigl(\mathtt{fail\_prob}(s{-}1,\, p{-}1) + \mathtt{fail\_prob}(s{-}1,\, p{+}1)\bigr).
\end{aligned}
\end{equation}
This definition is well-founded because the parameter $s$ decreases on each recursive call, so we can define it as a valid {\texttt{spec}} function in Verus.
The first case is when we successfully reach $0$ and terminate.
The second case is when we run out of fuel and have not yet reached $0$.
In that case, the probability that we will not terminate after $0$ more rounds is $1$. 
The third case captures how the random walk updates the position.

Since we have defined $\mathtt{fail\_prob}$ to be the amount of credit we need to ensure termination,
if we prove $ \forall \delta > 0.\;\exists s.\; \mathtt{fail\_prob}(s, \mathtt{pos}) < \delta$,
then this means that for any positive credit budget $\delta$, there is some $s$ such that we can use it to
pay for the program's termination within $s$ steps.
Proving this is a pure mathematical fact about the recurrence given by \eqref{eq:fail-prob}.
An explanation of this proof can be found in \autoref{sec:rw-convergence}.
In particular, in Verus, this is a straightforward fact we prove about the {\texttt{spec}} function $\mathtt{fail\_prob}$.

Once we have this fact, we turn to actually reasoning about the code implementing {\texttt{random\_walk}}.
We first use the \rref{Err-Thin-Air} rule to get some arbitrary $\err > 0$.
Using the property just proved for ${\mathtt{fail\_prob}}$, we get some fuel
\lstinline[language=Verus]|s| large enough that $\mathtt{fail\_prob}(\mathtt{s}, \mathtt{pos}) < \err$.
The remaining work is to show that, when we update our error credits using \ruleref{Err-Rand-Exp}, the resulting credits we use match the cases of the recurrence relation defining ${\mathtt{fail\_prob}}$.

The steps of the general pattern are to:
(1) define a pure mathematical recurrence relation for the probability of non-termination after $s$ steps, 
(2) prove that for any $\err > 0$ there exists $s$ large enough that $\err$ is greater than this probability of non-termination, and 
(3) generate a thin-air credit and prove a correspondence between the recurrence relation and the code.
Isolating the pure mathematical reasoning in steps 1 and 2 from the code reasoning in step 3 helps to structure the proofs.
In \Cref{sec:sampler-correctness}, we will see how to apply this pattern to several challenging case studies.

We have verified several other examples from \citet{eris}, including almost-sure termination
of the escaping spline \cite{McIverMKK18} and a higher-order rejection sampler.
In fact, it is no accident that this approach to reasoning about almost-sure termination works so well.
Recent work~\cite{completeness} establishes that the Eris program logic is \emph{complete} for almost-sure termination on a
higher-order probabilistic language: any program that terminates with probability $1$ can,
in principle, be proven to do so using error credits by picking the right credit distribution.\footnote{\citet{completeness} restrict to programs without dynamic allocation because Eris's rules are incomplete for reasoning about the memory addresses an allocator will assign, but this is orthogonal to reasoning about termination probabilities.}
From a practical standpoint, this gives
us confidence that the almost-sure termination support we expose in Verus is not artificially limited to a
narrow class of programs.
Although the hard part of solving the recurrence relation remains, the rules are expressive enough to cover a wide class of programs.

\section{Verifying Sampler Correctness}
\label{sec:sampler-correctness}

Now that we have seen how \thetool's encoding of error credits works for verifying some simple bounds and proving almost-sure termination,
we turn to proving the correctness of samplers written in Rust.
As explained in \Cref{par:sampler-correctness}, \citet{cpp26-distribution} proved a soundness theorem for Eris, which showed that, by proving a specification about a program $\expr$ that allows for error credits to be distributed in a way that preserves expected values under a distribution $\mu$, we can establish that $\expr$ in fact returns samples according to $\mu$. We call specifications of this form \rref{EPT} in the rest of this section.
Using such specifications, we prove the correctness of a number of samplers,
including the discrete Laplace and discrete Gaussian samplers (\Cref{sec:cks}),
the alias method (\Cref{sec:alias}), and the fast loaded dice roller (\Cref{sec:fldr}).

To illustrate the pattern, we warm up with a simple example: a sampler for the Bernoulli distribution 
$\mathrm{Bern}(p)$, where $p$ is a rational number of the form $a/b$. The sampler draws a
single uniform value $u$ over $\{0, \dots, b-1\}$ and returns whether $u$ falls below the numerator
$a$, which happens with probability exactly $a/b = p$.
\begin{lstlisting}[language=Verus,style=VerusLineNos,columns=fixed,literate={}]
fn sample_bernoulli_rational(a: &UBig, b: &UBig) -> bool {
    let u = rand_ubig(b);   // u ~ Uniform([0, b))
    u < a                   // true with probability a/b
}
\end{lstlisting}
Proving this sampler correct amounts to proving the following expectation-preserving specification,
in the form of \rref{EPT}, for every caller-supplied credit
allocation $\Err$ over the two outcomes:
\[
  \inferrule
    {\err \;\ge\; p\,\Err(\texttt{true}) + (1 - p)\,\Err(\texttt{false})}
    {\proves \thoare{\upto{\err}}{\texttt{sample\_bernoulli\_rational}(a, b)}{v.\; \upto{\Err(v)}}}
\]
Concretely, to prove this specification, we are given an $\Err$ satisfying the inequality in the premise, and a precondition of $\upto{\err}$ error credits.
We must figure out, based on $\Err$, what credit allocation function $f$ to supply to the call of \lstinline[language=Verus]|rand_ubig|.
To do so, we essentially perform backwards reasoning, determining what values of \lstinline[language=Verus]|u| cause \lstinline[language=Verus]|sample_bernoulli_rational| to return \texttt{true} and \texttt{false}.
From this, we get the requirement that $f(\texttt{u}) = \Err{(\texttt{true})}$ for $\texttt{u} < a$ and $f(\texttt{u}) = \Err{(\texttt{false})}$ for $\texttt{u} \ge a$.
We then check that this choice of $f$ satisfies the precondition for \lstinline[language=Verus]|rand_ubig|, which follows since
\begin{equation}\label{eq:bern-check}
  \tfrac1b\textstyle\sum_{u = 0}^{b-1} f(u)
  \;=\; \tfrac1b\bigl(a\,\Err(\texttt{true}) + (b - a)\,\Err(\texttt{false})\bigr)
  \;=\; p\,\Err(\texttt{true}) + (1 - p)\,\Err(\texttt{false})
  \;\leq\; \err.
\end{equation}
The postcondition of
\lstinline[language=Verus]|rand_ubig| then gives us the appropriate number of credits to prove the postcondition.

\paragraph{General Proof Strategy.}
The general recipe for proving sampler correctness is to do a form of backwards reasoning.
Letting $\mathit{ret}$ be the final return value of the sampler function, we trace backwards to find a symbolic expression for
$\Err(\mathit{ret})$ in terms of the intermediate random samples the code generates to compute $\mathit{ret}$.
This formula determines the shape of the 
credit allocation for each intermediate sampler call as we work backwards.
Tracing back to the beginning of the sampler routine, we prove a mathematical fact that the overall symbolic formula
is bounded by the expectation of $\Err$ under the distribution that the sampler is intended to generate.

If the sampler involves loops or recursion, the symbolic formula for the credit allocation will naturally be structured as a recurrence
relation, which we solve or bound by induction.
For these samplers we also have to prove termination using a \texttt{decreases} clause. 
To handle this, we use a \emph{separate} supply of error credits, generated from an initial thin-air credit and amplified using the approach
described in \Cref{sec:ast}.
This separates out the termination reasoning from reasoning about whether each sample has the correct weight.

In addition, we structure our Verus proofs so that all of the pure mathematical reasoning, such as
bounding a recurrence relation or proving that the symbolic formula is algebraically equivalent to the appropriate
expected value, is isolated to separate lemmas. 
The core reasoning about the executable code then just reduces to proving that the symbolic formula accurately reflects
the credit transformations needed at intermediate sampler calls.
A benefit of this decomposition is that the pure mathematical facts are easily discharged by a combination of Verus's SMT automation and LLM agents.
In the following examples, all of the Verus proofs for intermediate mathematical lemmas were done by Claude Opus 4.7 and 4.8, without human intervention.

\subsection{Discrete Gaussian}
\label{sec:cks}

\begin{wrapfigure}{r}{0.40\textwidth}
\centering
\begin{tikzpicture}[
  line width=0.3pt,
  samp/.style={draw, rounded corners, align=center, font=\scriptsize, inner sep=2pt, minimum width=3.6cm},
  dep/.style={->},
]
  \node[samp] (ru) at (0, 0.00) {\texttt{rand\_ubig}~(\S\ref{sec:axioms})};
  \node[samp] (br) at (0,-0.55) {\texttt{sample\_bernoulli\_rational}~(\S\ref{sec:sampler-correctness})};
  \node[samp] (e1) at (0,-1.10) {\texttt{sample\_bernoulli\_exp1}~(\S\ref{sec:cks-bern-exp1})};
  \node[samp] (e)  at (0,-1.65) {\texttt{sample\_bernoulli\_exp}~(\S\ref{sec:cks-bern-exp1})};
  \node[samp] (gs) at (0,-2.20) {\texttt{sample\_geometric\_exp\_slow}~(\S\ref{sec:geometric-bernoulli})};
  \node[samp] (gf) at (0,-2.75) {\texttt{sample\_geometric\_exp\_fast}~(\S\ref{sec:geometric-bernoulli})};
  \node[samp] (l)  at (0,-3.30) {\texttt{sample\_discrete\_laplace}~(\S\ref{sec:discrete-laplace})};
  \node[samp] (g)  at (0,-3.85) {\texttt{sample\_discrete\_gaussian}~(\S\ref{sec:discrete-gaussian})};
  \draw[->] (ru) -- (br);
  \draw[->] (br) -- (e1);
  \draw[->] (e1) -- (e);
  \draw[->] (e)  -- (gs);
  \draw[->] (gs) -- (gf);
  \draw[->] (gf) -- (l);
  \draw[->] (l)  -- (g);
  \draw[dep] (ru.east) to[out=0,in=0,looseness=0.4] (gf.east);
  \draw[dep] (e.east)  to[out=0,in=0,looseness=0.4] (g.east);
  \draw[dep] (br.west) to[out=180,in=180,looseness=0.4] (l.west);
\end{tikzpicture}
\caption{Call graph of the CKS algorithm for the discrete Gaussian: an arrow $A \to B$ means $A$ is used as a
sub-sampler by $B$.}
\label{fig:gaussian-callstack}
\end{wrapfigure}
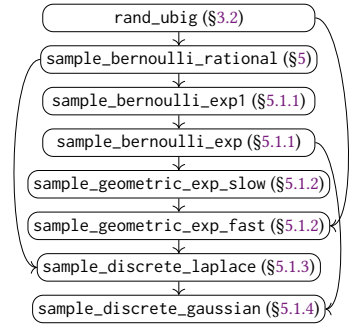
We prove the correctness of the discrete Laplace and discrete Gaussian samplers implemented
in OpenDP~\cite{opendp}, which use the CKS algorithm~\cite{cks16}.

The samplers are verified in three phases:
first, we verify a sampler for $\mathrm{Bern}(e^{-x})$ for $x \ge 0$;
then we use it to build two different samplers for $\mathrm{Geom}(1 - e^{-x})$;
finally, we use the geometric sampler to build the discrete Laplace and discrete Gaussian samplers.
All of these algorithms rely only on the single primitive sampler {\texttt{rand\_ubig}} we axiomatized 
in \Cref{sec:axioms}.
Each sampler is verified against an \rref{EPT} specification and reused as a sub-sampler by the
next, forming the call graph in \Cref{fig:gaussian-callstack}.

\subsubsection{Negative Exponential Bernoulli}
\label{sec:cks-bern-exp1}
We start with a sampler for the negative exponential Bernoulli distribution $\text{Bern}(e^{-x})$.
We represent $x$ as a rational number with \texttt{RBig} in Rust, which is an arbitrary rational number 
type, which can be broken down into the numerator and denominator as \texttt{IBig}/\texttt{UBig}.
First, we implement \texttt{sample\_bernoulli\_exp1}, which imposes the restriction that $x \in [0, 1]$.
This is then called by \texttt{sample\_bernoulli\_exp}, which drops the upper bound restriction.

\begin{wrapfigure}{r}{0.42\textwidth}
\begin{lstlisting}[language=Verus,style=VerusLineNos]
fn sample_bernoulli_exp1(x: RBig) -> bool {
    let mut k = UBig::ONE;
    loop {
        if sample_bernoulli_rational(&x, &k) { 
            k += UBig::ONE; 
        } else { return k.is_odd(); }
    }
}
\end{lstlisting}
\label{fig:bernoulli-exp1}
\end{wrapfigure}
The sampler flips $\mathrm{Bern}(x/k)$ for increasing $k = 1, 2, \dots$, and as soon as a flip returns false, it returns whether the current $k$ is odd.
In order to show that this code samples from $\text{Bern}(e^{-x})$, it suffices to prove an \rref{EPT} specification of the following form,
where $\Err$ is the caller-supplied credit allocation.
\begin{flalign*}
  & \inferrule
    {\err \;\ge\; e^{-x}\,\Err(\texttt{true}) + (1 - e^{-x})\,\Err(\texttt{false})}
    {\proves \thoare{\upto{\err}}{\texttt{sample\_bernoulli\_exp1}(x)}{b.\; \upto{\Err(b)}}} &
\end{flalign*}

We prove this specification by reasoning backwards from the credit allocation $\Err$ to construct the credit allocation we hand to each
sub-sampler, and then discharge that sub-sampler's \rref{EPT} rule.
The one new challenge is that
\texttt{sample\_bernoulli\_exp1} loops rather than making a single draw, so we have to maintain 
a loop invariant to save enough credits for each draw. 
This invariant keeps track of the credit $\err_k$ on entry to the $k$th iteration, ensuring it suffices
to cover the conditional expectation of $\Err$ over the eventual result,
\[
  \err_k \;\ge\; p_k\,\Err(\texttt{true}) + (1-p_k)\,\Err(\texttt{false}),
\]
where $p_k$ is the probability that the sampler eventually returns \texttt{true} when the loop has
reached iteration $k$. We get the following recurrence by conditioning on the outcome of the step-$k$ flip $\mathrm{Bern}(x/k)$, which continues to
step $k+1$ with probability $x/k$ and otherwise stops and returns whether $k$ is odd.
\[
  p_k \;=\; \tfrac{x}{k}\,p_{k+1} + \bigl(1-\tfrac{x}{k}\bigr)\,[k\text{ odd}],
  \qquad p_1 = e^{-x}.
\]
Here, $[k\text{ odd}]$ denotes the Iverson bracket, which is $1$ when $k$ is odd and $0$ otherwise.
At iteration $1$ the invariant is exactly the caller's precondition, since $p_1 = e^{-x}$.
The only analytic fact the argument needs about the $p_k$ is that each is a genuine probability in $[0,1]$,
which follows from an alternating Taylor-series bound on $e^{-x}$.

Reasoning backwards through the step-$k$ call to \texttt{sample\_bernoulli\_rational}
tells us the credit allocation function must be
\[
  b \;\mapsto\;
  \begin{cases}
    \err_{k+1}               & b = \texttt{true}\ \ (\text{continue}),\\
    \Err([k\text{ odd}]) & b = \texttt{false}\ \ (\text{return } \texttt{k.is\_odd()}),
  \end{cases}
  \qquad
  \err_{k+1} \;=\; \tfrac{k}{x}\,\err_k - \bigl(\tfrac{k}{x}-1\bigr)\Err([k\text{ odd}]),
\]
where $\err_{k+1}$ is solved for precisely so that \texttt{sample\_bernoulli\_rational}'s \rref{EPT} holds. The rest of the proof checks that the credit allocation preserves the loop invariant at $k+1$, which is a rearrangement of terms. We omit this here; further details can be found in \autoref{sec:bern-exp1-invariant}.

\paragraph{Modeling $e^{-x}$.} Z3 does not natively support transcendental 
functions like the natural exponential. We therefore expose $e^{(-)}$ as an
uninterpreted function in Verus and axiomatize the properties we need. The key Taylor-tail
bound used above is the only nontrivial axiom, and is described in \autoref{sec:taylor-axiom}; we verify it separately in Lean.

\begin{wrapfigure}{r}{0.45\textwidth}
\begin{lstlisting}[language=Verus,style=VerusLineNos]
fn sample_bernoulli_exp(x: &RBig) -> bool {
    let mut whole = x.floor();
    while whole > IBig::ZERO {
        if !sample_bernoulli_exp1(RBig::ONE) {
            return false;
        }
        whole -= IBig::ONE;
    }
    sample_bernoulli_exp1(x.frac())
}
\end{lstlisting}
\label{fig:bernoulli-exp}
\end{wrapfigure}
\paragraph{Generalizing to $\mathrm{Bern}(e^{-x})$ for $x \ge 0$.}
We next extend the sampler to arbitrary nonnegative $x$ using \texttt{sample\_bernoulli\_exp}, which calls
the previous \texttt{sample\_bernoulli\_exp1}. This sampler performs $\lfloor x \rfloor$ independent $\mathrm{Bern}(e^{-1})$ flips, returning false as soon as any of them fails, and if all succeed, it returns a final $\mathrm{Bern}(e^{-\operatorname{frac}(x)})$ flip.
All of these flips are independent, so the probability they all return true is given by the product of the probabilities that each individually returns true.
This is $e^{-x}$, since  $e^{-x} = e^{-\lfloor x \rfloor} \cdot e^{-(x - \lfloor x \rfloor)}$.

Working backwards, we find that the credit allocation function supplied to the final call to \texttt{sample\_bernoulli\_exp1(x.frac())}
is exactly $\Err$. For the loop,
we establish a similar loop invariant that the held credit dominates the conditional expectation of $\Err$.
Writing $r$ for the value of \texttt{x} still remaining, this invariant says that we have $\err$ error credits with 
\[
  \err \;\ge\; e^{-r}\,\Err(\texttt{true}) + (1-e^{-r})\,\Err(\texttt{false}).
\]
Concretely, to the iteration's \texttt{sample\_bernoulli\_exp1(1)} call, we give 
the credit allocation $F$:
\[
  \mathit{F}(b) \;=\;
  \begin{cases}
    e^{-(r-1)}\,\Err(\texttt{true}) + (1-e^{-(r-1)})\,\Err(\texttt{false})
      & b = \texttt{true}\ \ (\text{continue},\ r \mapsto r-1),\\
    \Err(\texttt{false}) & b = \texttt{false}\ \ (\text{return false}).
  \end{cases}
\]
On \texttt{true} it forwards the invariant's credit for the smaller problem $e^{-(r-1)}$, and on
\texttt{false} it pays $\Err(\texttt{false})$.
The rest is to check that the \rref{EPT} for \texttt{sample\_bernoulli\_exp1(1)} preserves the 
loop invariant:
\[
  e^{-1}\,\mathit{F}(\texttt{true}) + (1-e^{-1})\,\mathit{F}(\texttt{false})
  \;=\; e^{-r}\,\Err(\texttt{true}) + (1-e^{-r})\,\Err(\texttt{false}).
\]

\subsubsection{Geometric Bernoulli}
\label{sec:geometric-bernoulli}

We build a $\mathrm{Geom}(1 - e^{-x})$ sampler from the negative exponential Bernoulli sampler, and it has
a fast and a slow version.
The slow version is a standard geometric sampler, similar to the one in \Cref{sec:ast}.
It repeatedly draws $\mathrm{Bern}(e^{-x})$ until it returns false, and returns the number of draws that returned true. The fast version calls the slow version as a sub-sampler.
We omit the verification of the slow version and focus on the fast one.

Write $x = n/d$ in lowest terms, with $n$ and $d$ positive integers.
The sampler draws two independent samples and combines them by
integer division. The exponential rejection sampler (\texttt{sample\_exp\_rejection}) repeatedly
draws $u$ uniformly from $\{0,\dots,d-1\}$ and keeps it with probability $e^{-u/d}$, so it returns
$u$ with probability $e^{-u/d}/N$, where $N = \sum_{u=0}^{d-1} e^{-u/d}$. Then it calls the ``slow"
geometric sampler $v \sim \mathrm{Geom}(1 - e^{-1})$, and returns $\lfloor (u + d\,v)/n \rfloor$.
\begin{center}
\begin{minipage}[t]{0.44\linewidth}
\begin{lstlisting}[language=Verus,style=VerusLineNos]
fn sample_exp_rejection(d: &UBig) -> UBig {
    loop {
        let u = sample_uniform_ubig_below(d);
        if sample_bernoulli_exp_ubig(&u, d) {
            return u;
        }
    }
}
\end{lstlisting}
\end{minipage}\hfill
\begin{minipage}[t]{0.54\linewidth}
\begin{lstlisting}[language=Verus,style=VerusLineNos]
pub fn sample_geometric_exp_fast(x: RBig) -> UBig {
    let ((_, n), d) = x.into_parts(); // x = n/d
    let u = sample_exp_rejection(&d);
    let v = sample_geometric_exp_slow(&RBig::ONE);
    (v * d + u) / n
}
\end{lstlisting}
\end{minipage}
\end{center}

Aside from the integer floor divisions, the structure here is straightforward, and the credit reasoning is mechanical.

We first prove the exponential-rejection sampler's \rref{EPT} specification:

\[
  \inferrule
    {N = \textstyle\sum_{u=0}^{d-1} e^{-u/d} \\
     \err \;\ge\; \expect[u\sim\mathrm{rejection}]{\Err(u)}
       \;=\; \tfrac1N\textstyle\sum_{u=0}^{d-1} e^{-u/d}\,\Err(u)}
    {\proves \thoare{\upto{\err}}{\texttt{sample\_exp\_rejection}(d)}{u.\; \upto{\Err(u)}}}
\]
Since it's a stateless rejection sampler, the loop invariant preserves the $\err$
credits we started with,
as each iteration is independent.
This forces the allocation function for
$\mathrm{Bern}(e^{-u/d})$ to pay out $\Err(u)$ on \texttt{true} and carry 
$\err$ back to the next iteration on \texttt{false}.
And the credit allocation given to the uniform draw of $u$ is the $\mathrm{Bern}(e^{-u/d})$-average,
\[
  h(u) \;=\; e^{-u/d}\,\Err(u) + (1 - e^{-u/d})\,\err.
\]
Averaging over the uniform draw and using the precondition $\err \ge \tfrac1N\sum_{u=0}^{d-1} e^{-u/d}\,\Err(u)$,
\[
  \tfrac1d\sum_{u=0}^{d-1} h(u)
    \;=\; \tfrac{N}{d}\Bigl(\tfrac1N\sum_{u=0}^{d-1} e^{-u/d}\,\Err(u)\Bigr) + \bigl(1 - \tfrac{N}{d}\bigr)\err
    \;\le\; \err,
\]
so the held credit $\err$ is preserved across iterations. 
The almost sure termination is given by the $\mathsf{thin\_air}$ credit,
as each iteration amplifies it by a fixed factor
$\tfrac{1}{1 - N/d} = \tfrac{d}{d - N} > 1$.

The right-hand program is then straight-line, and the \rref{EPT} credit
allocations compose: working backwards, the slow geometric sampler is handed
$g(u,v) = \Err\big(\lfloor (u + d\,v)/n \rfloor\big)$, which forces the rejection sampler's allocation
$f(u) = \expect[v\sim\mathrm{Geom}(1-e^{-1})]{g(u,v)}$. The remaining proof obligation is then to check that the precondition has enough credits to cover the credit allocation for the exponential rejection sampler, namely $\sum_{r=0}^{\infty}(e^{-n/d})^r(1-e^{-n/d})\,\Err(r) \ge \expect[u\sim\mathrm{rejection}]{f(u)}$. The proof is a rearrangement of
two series, using a bijection between the terms, which we defer to \autoref{sec:fast-geometric}.

\subsubsection{Discrete Laplace}
\label{sec:discrete-laplace}
The discrete Laplace $\mathrm{Lap}(0,\mathit{scale})$ is symmetric around $0$: writing
$p = e^{-1/\mathit{scale}}$, it has probability mass function
$\mu_L(x) = \tfrac{1-p}{1+p}\,p^{|x|}$ for $x \in \mathbb{Z}$.
\footnote{
This looks different from the standard Laplace density:
$\tfrac{1}{2\sigma}\,e^{-|x|/\sigma}$ for $\sigma = \mathit{scale}$. Since $p = e^{-1/\sigma}$, $\mu_L(x) = \tfrac{e^{1/\sigma}-1}{e^{1/\sigma}+1}\,e^{-|x|/\sigma}$, with the same $e^{-|x|/\sigma}$, and the normalizer sums to $\sum_{x\in\mathbb{Z}} e^{-|x|/\sigma} = \tfrac{1+p}{1-p}$.
}

The sampler flips a fair coin for the sign and draws the magnitude from
$\mathrm{Geom}(1-p)$. It rejects when the magnitude is $0$ and the sign is negative so that we do not double-count $0$.

\begin{lstlisting}[language=Verus,style=VerusLineNos]
pub fn sample_discrete_laplace(scale: &RBig) -> IBig {
    loop {
        let positive = sample_bernoulli_rational(1, 2);                 // fair sign
        let k: IBig = sample_geometric_exp_fast(scale.recip()).into();  // |y| ~ Geom(1-p)
        if positive || !k.is_zero() {        // reject only the (-, 0) duplicate, then retry
            return if positive { k } else { -k };
        }
    }
}
\end{lstlisting}
This is essentially a rejection sampler wrapping the previous geometric sampler.
Thus, we can verify it
similarly to the exponential-rejection sampler (\Cref{sec:geometric-bernoulli}).
Reading the allocation off the program backwards, the loop maintains
$\err = \sum_{x\in\mathbb{Z}} \mu_L(x)\,\Err(x)$, and backward execution through the fair sign flip
$\mathrm{Bern}(\tfrac12)$ splits the held credit into a positive and a negative branch budget
$\err_{+}, \err_{-}$ (the negative branch carrying $\err$ back on the rejected $(-,0)$ outcome).
Discharging the sign flip's \rref{EPT} precondition then reduces to
$\err_{+} + \err_{-} \le 2\,\err$, which we verify in \autoref{sec:discrete-laplace-credit}.

\subsubsection{Discrete Gaussian}
\label{sec:discrete-gaussian}
The discrete Gaussian $\mathcal{N}_{\mathbb{Z}}(0,\sigma^2)$ is sampled by rejection against a
discrete-Laplace proposal: with $t = \lfloor\sigma\rfloor + 1$, we draw $y \sim \mathrm{Lap}(0,t)$ and
accept it with probability $e^{-\mathrm{bias}(y)}$, where $\mathrm{bias}(y) = (|y| - \sigma^2/t)^2/(2\sigma^2)$.
Its verification uses the same rejection-sampler technique as the discrete Laplace.
We defer the sampler and its correctness proof to \autoref{sec:discrete-gaussian-appendix}.

\subsection{Alias Method}
\label{sec:alias}

\begin{wrapfigure}{r}{0.28\textwidth}
\centering
\begin{tikzpicture}[x=1.0cm, y=0.14cm,
  lbl/.style={font=\small},
  bcap/.style={font=\footnotesize},
  c0/.style={fill=blue!18}, c1/.style={fill=red!18}, c2/.style={fill=green!22},
]
  \fill[c0] (0,0) rectangle (0.8,14);  \fill[c2] (0,14) rectangle (0.8,19);
  \draw (0,0) rectangle (0.8,19);
  \node[lbl] at (0.4,7) {$0$};  \node[lbl] at (0.4,16.5) {$2$};
  \fill[c1] (1,0) rectangle (1.8,12);  \fill[c0] (1,12) rectangle (1.8,19);
  \draw (1,0) rectangle (1.8,19);
  \node[lbl] at (1.4,6) {$1$};  \node[lbl] at (1.4,15.5) {$0$};
  \fill[c2] (2,0) rectangle (2.8,19);  \draw (2,0) rectangle (2.8,19);
  \node[lbl] at (2.4,9.5) {$2$};
  \node[bcap] at (-0.45,19) {$m$};  \node[bcap] at (-0.45,0) {$0$};
  \foreach \x/\t in {0/0, 1/1, 2/2} { \node[bcap] at (\x+0.4,-2.5) {bin~\t}; }
\end{tikzpicture}
\caption{The alias table for weights $(a_0,a_1,a_2) = (7,4,8)$.}
\label{fig:alias}
\end{wrapfigure}
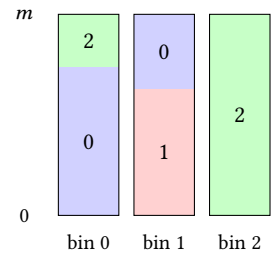
From the previous subsection, we see that \thetool can verify samplers with intricate 
probabilistic behavior.
In this subsection and the next, we turn to examples that combine probabilistic reasoning with data structures built from mutable state.
These demonstrate the benefit of \thetool's ability to build on Verus's existing support for reasoning about stateful data structures.

We start with the alias method
for sampling from finite discrete distributions~\cite{walker_alias,vose_alias,schwarz_blog}.
Given a finite set of $n$ labels $\{0, \dots, n-1\}$ and 
a vector of nonnegative integer weights $(a_0, a_1, \ldots, a_{n-1})$, 
the alias method samples from the distribution that returns label $i$ with 
probability $a_i / \sum_{j=0}^{n-1} a_j$.

The algorithm is split into two functions: a preprocessing function and a sampling function.
The preprocessing function returns an \emph{alias table} that is used for subsequent sampling.
The table consists of two length-$n$ arrays,
$\mathit{prob}$ and $\mathit{alias}$. Picture $n$ equal-probability \emph{bins}, one per index, each
filled with a total of $m$ units of mass. Bin $i$ is split between at most two labels: it holds
$\mathit{prob}[i]$ units of its own label $i$ and the remaining $m - \mathit{prob}[i]$ units of an
\emph{alias} label $\mathit{alias}[i]$. Summing over
all bins shows that each label $k$ owns exactly $n\,a_k$ units. 
\Cref{fig:alias} shows an alias table for the weights $(a_0, a_1, a_2) = (7, 4, 8)$.
Each bin has $m = 19$ units, with $\mathit{prob} = [14,12,19]$, $\mathit{alias} = [2,0,2]$.
For example, bin $0$ holds $\mathit{prob}[0] = 14$ units of label $0$ and the remaining $5$
units of label $2$.

Once the table has been constructed, the sampling routine starts by picking
a bin $i$ uniformly.
Then, it samples a second value from $\{0, \dots, m-1\}$, corresponding to a unit
in the bin, and returns the label for that unit.

Because the distribution of the samples returned by the sampling function depends on the preprocessed
table it is given as input, the specification of the sampler is \emph{parametric} over the
table's weights.
The table is represented by the executable \texttt{AliasTable}, and the
sampler's only assumption about it is a well-formedness predicate \lstinline[language=Verus]|wf|.

\noindent
\begin{minipage}[t]{0.48\linewidth}
\begin{lstlisting}[language=Verus,style=VerusLineNos]
pub struct AliasTable {
    pub n: u64,      // # of labels
    pub m: u64,      // total weight
    pub weights: Vec<u64>, // a_0..a_{n-1}
    pub prob: Vec<u64>,    // own units per bin
    pub alias: Vec<u64>,   // alias label per bin
}
\end{lstlisting}
\end{minipage}\hfill
\begin{minipage}[t]{0.48\linewidth}
\begin{lstlisting}[language=Verus,style=VerusLineNos]
pub struct Alias {   // ghost view of AliasTable
    pub n: nat,
    pub m: nat,
    pub weights: spec_fn(nat) -> nat,
    pub prob: spec_fn(nat) -> nat,
    pub alias: spec_fn(nat) -> nat,
}
\end{lstlisting}
\end{minipage}

Each executable \texttt{AliasTable} has a ghost view \lstinline[language=Verus]|self@ : Alias|, in which
the concrete \lstinline[language=Verus]|u64| fields become \lstinline[language=Verus]|nat| and the
\lstinline[language=Verus]|Vec<u64>| arrays become mathematical functions
\lstinline[language=Verus]|spec_fn(nat) -> nat|. The well-formedness predicate \lstinline[language=Verus]|wf|
requires that the vectors have length \lstinline[language=Verus]|n| and that the ghost view satisfies the
key condition \lstinline[language=Verus]|valid_alias(self@)|, a pure predicate on the \texttt{Alias}
view:

\begin{lstlisting}[language=Verus,style=VerusLineNos,columns=fixed,literate={}]
  pub open spec fn valid_alias(t: Alias) -> bool {
    &&& t.n >= 1
    &&& t.m >= 1
    &&& t.m == sum_of_weights(t, t.n)
    &&& (forall |i: nat| i < t.n ==> (t.prob)(i) <= t.m)
    &&& (forall |i: nat| i < t.n ==> (t.alias)(i) < t.n)
    // every label k's total units equal n*a_k  (Vose's redistribution invariant).
    &&& (forall |k: nat| k < t.n ==> label_units(t, t.n, k) == t.n * (t.weights)(k))
}
\end{lstlisting}
The first 5 conjuncts are nondegeneracy conditions, ensuring 
\lstinline[language=Verus]|AliasTable| has the right shape and that the $\mathit{prob}$ 
and $\mathit{alias}$ arrays are well-formed.
The last conjunct ensures that the relative weights are correct: for every label $k$,
$\mathit{label\_units}(t, n, k) = n\,a_k$, where $\mathit{label\_units}$ sums the units of label $k$
across all $n$ bins. This encodes that we have faithfully redistributed the weights into the bins, 
and is the key invariant that ensures the sampler returns label $k$ with probability $a_k/m$.
\noindent
\begin{minipage}[t]{0.48\linewidth}
\begin{lstlisting}[language=Verus,style=VerusLineNos,basicstyle=\ttfamily\tiny]
pub fn alias_preprocess(weights: Vec<u64>, m: u64) 
  -> (ret: AliasTable)
requires
    weights@.len() >= 1,
    m >= 1,
    (weights@.len() as nat) * (m as nat) <= u64::MAX as nat,
    seq_sum(weights@, weights@.len() as nat) == m as nat,
ensures
    wf(ret),
    ret.n as nat == weights@.len(),
    ret.m == m,
    ret.weights@ == weights@,
{
    /* body and proof omitted */
}
\end{lstlisting}
\end{minipage}\hfill
\begin{minipage}[t]{0.52\linewidth}
\begin{lstlisting}[language=Verus,style=VerusLineNos,basicstyle=\ttfamily\tiny]
pub fn sample_alias(
    tab: &AliasTable,
    Ghost(e): Ghost<spec_fn(nat) -> real>,
    Tracked(input_credit): Tracked<ErrorCreditResource>,
    Ghost(eps): Ghost<real>,
) -> ((value, out_credit): (u64, Tracked<ErrorCreditResource>))
    requires
        wf(tab),
        forall |x: nat| e(x) >= 0real,
        eps >= alias_exp(tab@, e),
        input_credit@ =~= Value { car: eps },
    ensures
        out_credit@@ =~= Value { car: e(value as nat) },
{
    let i = rand_u64(tab.n);   // bin  i ~ U{0..n}
    let r = rand_u64(tab.m);   // slot r ~ U{0..m}
    if r < tab.prob[i] { i }   // own-label part
    else { tab.alias[i] }      // alias part
}
\end{lstlisting}
\end{minipage}

The preprocessing function \texttt{alias\_preprocess} constructs the table and ensures it is well-formed in the postcondition.
Meanwhile, \texttt{sample\_alias} assumes well-formedness as a precondition.
In \texttt{sample\_alias}, the expected value of the user's credit allocation function \texttt{e} is computed with \texttt{alias\_exp}, which uses the distribution
represented by the table.

The credit allocation of the sampler itself is straightforward.
Conditioned on the drawn bin, the inner threshold draw is expectation-preserving
and returns $\upto{\mathcal{E}(\mathit{result})}$, so we fund the outer bin draw with its per-bin
average; regrouping those credits by label and applying the table's validity
($\mathit{label\_units}(t, n, k) = n\cdot\mathit{weights}(k)$) shows the sampler meets the precondition
$\err \ge \sum_{k<n}\tfrac{a_k}{m}\,\mathcal{E}(k)$.

Indeed, much of the challenge lies in showing that the preprocessing step constructs the table correctly.
We verify an $O(n)$ version~\cite{vose_alias} that manages two worklists to find labels that still have unaccounted-for units
during construction.
To the best of our knowledge, there is no prior work on verifying the alias method, as it involves
intricate reasoning between preprocessing and sampling.
This is precisely the modularity that error credits buy us.
The deterministic preprocessing step that builds a
well-formed table is verified with ordinary functional-correctness specifications,
while the probabilistic sampling step just updates credits, relying on the table being correctly
structured through the well-formedness assumption.

\subsection{Fast Loaded Dice Roller}
\label{sec:fldr}

Like the alias method, the fast loaded dice roller samples from finite discrete distributions through a
preprocessing stage that builds a table and a sampling stage that uses it. However,
unlike the alias method, it uses only a fair coin~\cite{fldr}.
The algorithm has two stages: a preprocessing
stage that constructs a table representing a discrete distribution generating (DDG) tree, and a
sampling stage that performs a random walk through this tree.
We first focus on a simpler version that only samples from a uniform distribution, called the
fast dice roller algorithm.

\subsubsection{Fast Dice Roller}

\begin{wrapfigure}{r}{0.38\textwidth}
\begin{lstlisting}[language=Verus,style=VerusLineNos,escapeinside={(*@}{@*)}]
pub fn sample_fdr(n: u64) -> u64 {
    let mut v: u64 = 1;              (*@\label{ln:fdr-initv}@*)
    let mut c: u64 = 0;              (*@\label{ln:fdr-initc}@*)
    loop {
        v = 2 * v;                  (*@\label{ln:fdr-double}@*)
        c = 2 * c + rand_2_u64();   (*@\label{ln:fdr-refine}@*)
        if v >= n {                  (*@\label{ln:fdr-test}@*)
            if c < n { return c }    (*@\label{ln:fdr-accept}@*)
            else { v = v - n;  c = c - n } (*@\label{ln:fdr-reject}@*)
        }
    }
}
\end{lstlisting}
\label{fig:fdr}
\end{wrapfigure}
The fast dice roller~\cite{DBLP:journals/corr/abs-1304-1916} samples uniformly from $\{0, 1, \dots, n-1\}$ using only fair coin flips.
It maintains a state $(v, c)$, where $v$ is the size of what is called the current \emph{window} and $c$ is uniform over
$\{0, 1, \dots, v-1\}$; it starts from the trivial window $v = 1$, $c = 0$. Each iteration
doubles the window and refines $c$ with one fresh coin bit
(lines~\ref{ln:fdr-double}-\ref{ln:fdr-refine}), which keeps $c$ uniform on
$\{0, \dots, v-1\}$. 
The window grows until $v \ge n$ (line~\ref{ln:fdr-test}), at which point $c$ is uniform
on a range of size $v \ge n$. If $c < n$, the value already lies in the target range and is returned
(line~\ref{ln:fdr-accept}), uniform on $\{0, \dots, n-1\}$ as required. Otherwise the algorithm
repeats the loop, shifting both down by $n$ (line~\ref{ln:fdr-reject}).

The correct credit allocation function to pass to the call to \texttt{rand\_2\_u64} depends
on the current state of $(v, c)$. 
Letting $\mathcal{E}$ be the user-supplied credit allocation function,
we define two mutually recursive functions, which are additionally indexed by a fuel parameter $k$.
\[
\begin{aligned}
  \mathrm{fdr}_f(v,c,0) ={}& 0                                &\qquad\quad& &\qquad\quad& \text{(ran out of fuel $k$)}\\
  \mathrm{fdr}_f(v,c,k) ={}& \rlap{$\tfrac12\bigl(\mathrm{fdr}_h(2v,\,2c,\,k{-}1)
                             + \mathrm{fdr}_h(2v,\,2c{+}1,\,k{-}1)\bigr)$} &\qquad\quad& &\qquad\quad&\\[4pt]
  \mathrm{fdr}_h(v,c,k) ={}& \mathcal{E}(c)                    &\qquad\quad& \text{if } v \ge n,\ c < n   &\qquad\quad& \text{(accept)}\\
                        {}& \mathrm{fdr}_f(v{-}n,\,c{-}n,\,k)  &\qquad\quad& \text{if } v \ge n,\ c \ge n &\qquad\quad& \text{(reject, restart)}\\
                        {}& \mathrm{fdr}_f(v,\,c,\,k)          &\qquad\quad& \text{if } v < n             &\qquad\quad& \text{(continue doubling)}
\end{aligned}
\]
These definitions are well-founded since the $k$ parameter decreases by $1$ when going from $\mathrm{fdr}_f$ to $\mathrm{fdr}_h$. 
Here $\mathrm{fdr}_f(v,c,k)$ is the conditional expectation $\expect{\mathcal{E}(\mathit{out}) \mid (v,c)}$
of the caller-supplied allocation over the returned value, taken over the next $k$ coin flips from the
state $(v,c)$: it models the first \texttt{flip} step in the loop. 
$\mathrm{fdr}_h$ models the post-doubling test: 
mimicking the program structure, it pays out $\mathcal{E}(c)$ on the accepting branch 
($v \ge n$, $c < n$), restarts the doubling
on the rejecting branch with the shifted window ($v \ge n$, $c \ge n$), and otherwise keeps doubling
($v < n$).

The heavy lifting lies in showing the following pure mathematical fact on $\mathrm{fdr}_f$ at the initial state $(1,0)$: \[
  \mathrm{fdr}_f(1,0,k) \;\le\; \tfrac1n\textstyle\sum_{i<n}\mathcal{E}(i),
\]
where the sum on the right-hand side is the expected value of $\mathcal{E}$ under the uniform distribution, as required by the \rref{EPT}-style specification.
We defer this proof to \autoref{sec:fdr-bound}.

\subsubsection{Fast Loaded Dice Roller}

The fast loaded dice roller generalizes and builds upon the previous algorithm by
sampling from finite discrete distributions where the weights of the choices need not be uniform.
In particular, the algorithm is parameterized by a choice of integer weights
$a_0, a_1, \ldots, a_{n-1}$ with total $m = \sum_{i=0}^{n-1} a_i$, and the algorithm samples outcome $i$ with
probability $a_i/m$, using only fair coin flips.

The preprocessing step compiles these weights into a discrete distribution generating (DDG)
tree: a binary tree whose leaves are labelled by outcomes in $\{0,\ldots,n\}$, where $n$ is an extra label.
The sampler walks down the tree one fair coin
flip at a time, descending left on $0$ and right on $1$, tracking the current depth $c$ and node
position $d$, and stopping once it hits a leaf. 
Starting from the root, the algorithm reaches a leaf at depth $c$ with probability $2^{-c}$.
If it reaches a leaf labelled with a value in $\{0, \dots, n-1\}$, it returns that value. Otherwise, on reaching a leaf with label $n$, it returns to the root of the tree and repeats.

To construct the tree, the first step is to assign a weight to outcome $n$ which
causes the total weight to go from $m$ to the next power of two, which we do
by setting $\mathit{depth} = \lceil \log_2 m \rceil$ and making $a_n =
2^{\mathit{depth}} - m$.
We then add leaves and label them so that if we sum the probabilities of all
leaves labelled $i$, the total is equal to $a_i/2^{\mathit{depth}}$.
The tree is represented by a table, which records, 
for each level $c$, the number of leaves $h[c]$ at that level and
their labels $\mathit{lab}[c]$; 
by convention the $h[c]$ leaves occupy positions $0,\ldots,h[c]-1$ and
the internal nodes follow at positions ${\ge}\,h[c]$. To decide which labels are at level $c$,
the preprocessing step reads the binary expansions of the weights 
$a_i$ and adds a leaf for each set bit $2^{\mathit{depth}-c}$ of $a_i$.

\begin{figure}[h]
\centering
\begin{minipage}[c]{0.52\linewidth}
\centering
\resizebox{\linewidth}{!}{%
\begin{tikzpicture}[
  level distance=10mm,
  every node/.style={font=\small},
  inner/.style={circle,draw,fill=black,inner sep=1.7pt},
  leaf/.style={draw,rounded corners,inner sep=2.5pt,minimum size=5.5mm},
  level 1/.style={sibling distance=24mm},
  level 2/.style={sibling distance=14mm},
  level 3/.style={sibling distance=11mm},
  level 4/.style={sibling distance=9mm},
  level 5/.style={sibling distance=9mm},
  edge from parent/.style={draw,-},
  backedge/.style={draw,->,dashed,black},
]
\node[inner] (root) {}
  child {node[inner] {}
    child {node[leaf] {$2$}}
    child {node[inner] {}
      child {node[leaf] {$0$}}
      child {node[leaf] {$1$}}
    }
  }
  child {node[inner] {}
    child {node[leaf] (rA) {$\mathsf{r}$}}
    child {node[inner] {}
      child {node[leaf] (rB) {$\mathsf{r}$}}
      child {node[inner] {}
        child {node[leaf] {$0$}}
        child {node[inner] {}
          child {node[leaf] {$0$}}
          child {node[leaf] (rC) {$\mathsf{r}$}}
        }
      }
    }
  };
\draw[backedge] (rA) to[out=180,in=270,looseness=1.1] (root.south);
\draw[backedge] (rB) to[out=180,in=250,looseness=1.0] (root.south west);
\draw[backedge] (rC) to[out=0,in=-35,looseness=0.8] (root.east);
\end{tikzpicture}%
}
\end{minipage}\hspace{0.5em}
\begin{minipage}[c]{0.22\linewidth}
\footnotesize
$\renewcommand{\arraystretch}{1.2}
\begin{array}{c|c|l}
  c & h[c] & \mathit{lab}[c]\\ \hline
  1 & 0 & [\,]\\
  2 & 2 & [\,2,\ \mathsf{r}\,]\\
  3 & 3 & [\,0,\ 1,\ \mathsf{r}\,]\\
  4 & 1 & [\,0\,]\\
  5 & 2 & [\,0,\ \mathsf{r}\,]
\end{array}$
\end{minipage}
\caption{The DDG tree (left) and the stored table $(h, \mathit{lab})$ (right) for the weights
$(7, 4, 8)$, the reject label $\mathsf{r} = 3$. Dashed
arrows show that reaching a reject leaf restarts the walk at the root.}
\label{fig:ddg}
\end{figure}
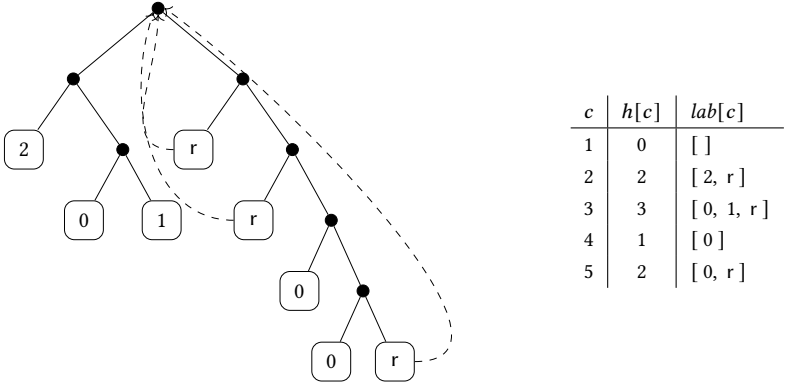

As a running example, take the weights $(a_0, a_1, a_2) = (7, 4, 8)$, so $m = 19$. Padding to the
next power of two adds a reject outcome $\mathsf{r}$ of weight $a_3 = 2^{\lceil \log_2 19 \rceil} - 19
= 32 - 19 = 13$, giving a total of $2^5 = 32$. Reading the binary expansions
$7 = (00111)_2$, $4 = (00100)_2$, $8 = (01000)_2$, and $13 = (01101)_2$, outcome $i$ receives one leaf
at depth $d$ for each set bit $2^{5-d}$ of $a_i$; a leaf at depth $d$ is reached with probability
$2^{-d}$, so the leaves of outcome $i$ contribute exactly $a_i / 32$. For instance, label $0$ has
leaves at depths $3, 4, 5$, contributing $\tfrac{4+2+1}{32} = \tfrac{7}{32}$. \autoref{fig:ddg} shows the resulting tree
and the stored table $(h, \mathit{lab})$, which records, for each level $c$, the number of leaves $h[c]$
and their labels $\mathit{lab}[c]$ in ascending order.

Similar to the alias method, the proof of the sampler is \emph{parametric} over 
the preprocessed data structure it is handed: the table is represented by the executable \texttt{FldrTable}, 
and the sampler's only assumption about it is a well-formedness predicate \lstinline[language=Verus]|wf|.
This predicate collects the pure properties the
table must satisfy for a faithful encoding of the DDG: the \lstinline[language=Verus]|Vec| fields have
the expected lengths, and the ghost view \lstinline[language=Verus]|t@| satisfies
\lstinline[language=Verus]|valid_ddg|. Just like \lstinline[language=Verus]|valid_alias|,
\lstinline[language=Verus]|valid_ddg| is a pure predicate on the ghost view of the table (whose
\lstinline[language=Verus]|Vec| fields become mathematical \lstinline[language=Verus]|spec_fn|s), but
its definition is considerably more involved, so we elide it here.

\noindent
\begin{minipage}[t]{0.48\linewidth}
\begin{lstlisting}[language=Verus,style=VerusLineNos]
pub struct FldrTable {
    pub n: u64,      // # of labels
    pub m: u64,      // total weight
    pub levels: u64, // ceil(log2 m)
    pub weights: Vec<u64>,  // a_0..a_{n-1}
    pub h: Vec<u64>,        // # leaves per level
    pub lab: Vec<Vec<u64>>, // labels in each level
}
\end{lstlisting}
\end{minipage}\hfill
\begin{minipage}[t]{0.50\linewidth}
\begin{lstlisting}[language=Verus,style=VerusLineNos]
pub open spec fn wf(t: FldrTable) -> bool {
  &&& valid_ddg(t@)
  &&& t.h@.len() == t.levels + 1
  &&& t.lab@.len() == t.levels + 1
  &&& forall|c| 0 <= c <= t.levels
        ==> t.lab@[c]@.len() == t.h@[c]
}
\end{lstlisting}
\end{minipage}

\noindent
\begin{minipage}[t]{0.48\linewidth}
\begin{lstlisting}[language=Verus,style=VerusLineNos]
pub fn fldr_preprocess(
    weights: Vec<u64>, m: u64, levels: u64,
) -> (tab: FldrTable)
    requires
        1 <= levels <= 62,
        pow2(levels as nat) <= usize::MAX as nat,
        1 <= m as nat <= pow2(levels as nat),
        /* ...more requires omitted... */
    ensures
        wf(tab),
        tab.n as nat == weights@.len(),
        tab.m == m,
        tab.levels == levels,
        tab.weights@ == weights@,
{
    /* body and proof omitted */
}
\end{lstlisting}
\end{minipage}\hfill
\begin{minipage}[t]{0.50\linewidth}
\begin{lstlisting}[language=Verus,style=VerusLineNos]
pub fn sample_fldr(
    tab: &FldrTable,
    Ghost(e): Ghost<spec_fn(nat) -> real>,
    Tracked(in_credit):Tracked<ErrorCreditResource>,
    Ghost(eps): Ghost<real>,
) -> ((val, out_credit):
      (u64, Tracked<ErrorCreditResource>))
    requires
        wf(tab),
        forall |x: nat| e(x) >= 0real,
        eps >= fldr_exp(tab@, e),
        in_credit@ =~= Value { car: eps },
    ensures
        out_credit@@ =~= Value { car: e(val as nat) }
{
    /* body and proof omitted */
}
\end{lstlisting}
\end{minipage}

The sampling routine takes such a table as input, and the distribution of the samples it returns depends on that table.
Again, the \rref{EPT} specification is therefore stated in terms of the distribution encoded by the table, where we use the helper function \texttt{fldr\_exp} to compute the expected value of the credit allocation function \texttt{e} based on this table.
The verification of the sampler is very similar to that of the fast dice roller, except that the credit
allocation now follows the tree's level/position structure rather than the doubling window. The
credit allocation and the sampler proof are in \autoref{sec:fdr-bound}.

\section{Extending VerusBelt for Probability}
\label{sec:verisbelt}

\newcommand{\fupdfin}[1][]{\pvs[#1]^{\mkern0.5mu\mathsf{fin}}}

The encoding of error credits in \thetool{} described in \Cref{sec:verus} only involves adding two trusted axioms, which are close in formulation to the rules in Eris.
Nevertheless, one might worry about potential unsoundness because the language considered in Eris is quite different from Rust, and some of Verus's other features have no analogue in Eris.
To address this and gain additional confidence in our encoding, this section adapts VerusBelt~\citep{verusbelt}, which was developed to justify the soundness of some of Verus's features.
Specifically, VerusBelt provides a semantic foundation for Verus's proof-oriented extensions to the
Rust type system.
It provides justification for a number of axioms in Verus, such as those for \texttt{PCell} and \texttt{PPtr}.
Just like the axioms that we add in \thetool{} to connect error credits to primitive sampling functions, these connect the tracked permissions for pointers to the actual functions that modify them.
Thus, by adapting the model of VerusBelt, we can similarly justify these new axioms for error credits.

\begin{figure}[h]
\centering
\begin{tikzpicture}[
  layer/.style={draw, rounded corners=2pt, minimum width=6.2cm, minimum height=0.95cm,
                align=center, font=\footnotesize, inner sep=2pt},
  foot/.style={font=\small},
]
  \node[layer]                    (Lbase)  at (-3.7, 0)    {Iris Base Logic};
  \node[layer, fill=RedDevil!10]  (Lwp)    at (-3.7, 0.95) {Iris WP};
  \node[layer]      (Lrb)    at (-3.7, 1.90) {RustBelt};
  \node[layer]      (Lpot)   at (-3.7, 2.85) {VerusBelt Proof-Oriented Types};

  \node[layer]                    (Rbase)  at (3.7, 0)    {Iris Base Logic};
  \node[layer, fill=DarkGreen!12] (Rwp)    at (3.7, 0.95) {Eris WP};
  \node[layer]      (Rrb)    at (3.7, 1.90) {RustBelt};
  \node[layer]      (Rpot)   at (3.7, 2.85) {VerusBelt Proof-Oriented Types};
  \node[layer, fill=DarkGreen!18] (Rec)    at (3.7, 3.80) {Error Credits};

  \draw[-{Stealth[length=2mm]}, thick]
    (Lwp.east) -- node[font=\scriptsize, above]{swap}
                  node[font=\scriptsize, below]{$+\;\upto{\err}$} (Rwp.west);
\end{tikzpicture}
\caption{The VerusBelt Proof-Oriented type proofs are written against a generic Iris
weakest precondition, so porting VerusBelt to a probabilistic model involves swapping
the definitions from Iris's \texttt{wp} to that of Eris.
On top, we additionally define the types for the error credit resources.
By ensuring that the Eris \texttt{wp} exposes the same required interface as the Iris version, we can preserve
most of the existing proofs from VerusBelt.}
\label{fig:wp-swap}
\end{figure}
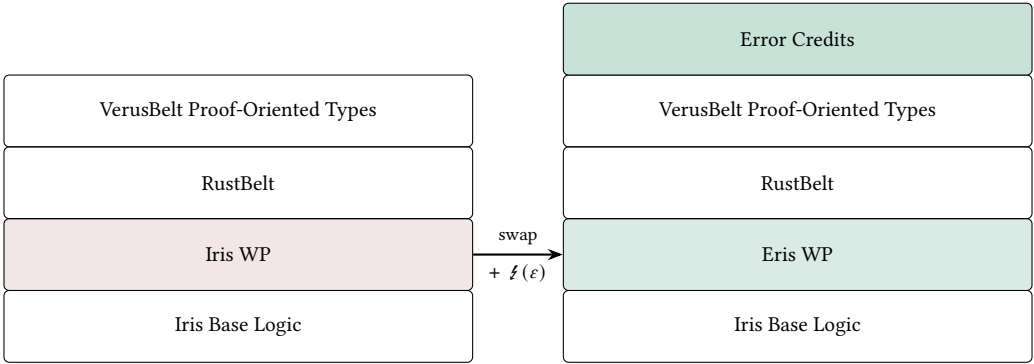

\newcommand\lambdaRust{\texorpdfstring{\ensuremath{\lambda_{\mathsf{Rust}}}}{lambdaRust}}

\Cref{fig:wp-swap} illustrates how VerusBelt is constructed in layers, and how our AlerusBelt model adapts them.
At the base of VerusBelt is the Iris base logic: this is the foundational assertion logic with separation logic connectives that Iris provides.
On top of this, Iris implements a language-generic weakest precondition assertion $\texttt{wp}$. 
This generic weakest precondition is instantiated for \lambdaRust{}, the model of a core subset of Rust used in RustBelt.
Above this, a logical-relations model of Rust's type system is constructed, in which types are interpreted as separation logic assertions in this instantiation of the weakest precondition.
Finally, VerusBelt itself extends this logical relations model with support for $\texttt{ensures}$ and $\texttt{requires}$ clauses, and validates the axioms for various primitives.

RustBelt and VerusBelt are substantial and complex Rocq developments.
Thus, to make this adaptation feasible, AlerusBelt ports the development in a way that largely avoids modifying these layers.
As shown in the figure, we replace the Iris weakest precondition with the Eris weakest precondition in the lower layers.
Since Eris is related to Iris, it is possible to do this replacement in a way that preserves the interface that RustBelt and VerusBelt need without disruption.
Then, on top of VerusBelt's existing constructions, we additionally model the error-credit resource and axioms we use, via a simple translation to the underlying error-credit rules that Eris provides.

However, pulling off the swap of Iris for Eris in this construction requires a few changes, since the formulation of Eris developed by \citet{eris} has certain mismatches with the standard Iris weakest precondition.
The rest of this section describes these issues and how we resolved them.
Several of these proof porting efforts were assisted by Opus 4.8. 

\paragraph{Later Credits}
VerusBelt uses an Iris feature called later credits~\citep{later-credits}, which helps manage the \emph{later modality} that occurs in Iris in order to soundly incorporate various forms of impredicativity.
However, Eris did not support later credits, as doing so posed an obstacle in its soundness proof.
The issue is that, under the hood, Iris's and Eris's weakest preconditions are defined in terms of various sequences of modalities that are used to
model features like updating ghost state, accessing Iris's impredicative invariants, and spending/redistributing error credits.
The soundness proof for Eris relies on various \emph{commutative laws} that allow for swapping the order of these modalities with other connectives,
in particular universal quantification.
Once later credits are included, however, these commutative laws are lost.
As a result, the existing Eris soundness proof does not work.

Recently, however, \citet{hupd} proposed an alternative formulation of several modalities in Iris and restructured Iris's soundness proof in a way that retains the key commutative laws, which was subsequently reworked by \citet{krebbers-later-credits-mr}.
By using their approach with these new modalities, we were able to add later credits into Eris and still derive soundness.

\paragraph{Prophecy Variables}
Since the work of \citet{jung-prophecy}, Iris has supported \emph{prophecy variables}, a mechanism that allows for ``predicting'' in the course of a proof what the future outcomes of various operations will be.
VerusBelt uses this to model an extension to Verus that supports more flexible reasoning about mutable references, following an approach introduced by \citet{rusthornbelt}.
However, Iris's prophecy variables are known to be unsound when combined with Eris's error credits~\citep{eris-noproph}.
Thus, we remove all features of VerusBelt that depend upon prophecy variables.
Fortunately, the only use is to model this recently implemented extension for reasoning about returning mutable references, and none of our example proofs make use of this feature.

\paragraph{Limitations}
Besides the removal of prophecy variables, AlerusBelt has two main limitations.
First, the original Eris does not support concurrency, and so we have additionally removed concurrency from \lambdaRust{} in porting the model.
However, none of our case studies involve concurrency.
A subsequent extension to Eris developed by \citet{coneris} does add concurrency support, and it would be interesting to port AlerusBelt to this logic to recover concurrency.

A second limitation arises from the fact that VerusBelt uses only a partial-correctness variant of Iris's weakest precondition and does not model Verus's \texttt{decreases} clauses for termination.
AlerusBelt thus similarly uses the partial-correctness version of Eris.
Thus, while our examples use total Eris's approach to proving almost-sure termination by credit amplification, the soundness of this aspect of our encoding is not captured in AlerusBelt.

Nevertheless, this model does provide some assurance that \thetool's encoding of error credits is compatible with Verus's other features.
This fact is \emph{a priori} non-obvious, as the example of prophecies for mutable borrows and the role of later credits in VerusBelt demonstrate.

\section{Related Work}

\paragraph{Semi-Automated Verifiers for Probabilistic Programs}

Caesar~\citep{caesar} is a deductive verifier for probabilistic programs that uses the HeyVL~\citep{heyvl} intermediate verification language.
In particular, in the latter, verification conditions are quantitative properties about expected values.
HeyVL and hence Caesar can check properties that are not directly captured by specifications with \thetool{}, such as bounds on expected values of running times and properties like \emph{positive} almost-sure termination.
On the other hand, supporting quantitative verification conditions requires a different design and implementation of the verification stack, as compared to other automated deductive verifiers.
Currently, Caesar targets a front-end language that is simpler than Rust and lacks many of the challenging language features Rust has.

\citet{cohen-pps17} develops an approach to reasoning about probabilistic behaviors in deductive verifiers by encoding probabilistic quantities as ghost values that are updated in an expectation-preserving way, much as Eris updates error credits.
He uses these to state probabilistic invariants over a system's behavior.
However, to the best of our knowledge, his encoding does not have an analogue of the thin-air credit rule, which we use to give modular proofs of almost-sure termination.
In addition, his ghost values are not embedded in a separation logic, and so do not support the kind of splitting via separating conjunction that error credits enjoy.

\citet{dafny-vmc} develop a library for probabilistic verification in Dafny.
The library uses a monadic style in which programs consume streams of random bits, following an approach pioneered by \citet{hurd-thesis}.
They apply this approach to verify the negative exponential Bernoulli sampler discussed in \Cref{sec:cks-bern-exp1}.
In contrast, \thetool{} can reason directly about probabilistic programs that are not written in monadic style.

\citet{arnold2024-ma-thesis} applies Prusti~\citep{prusti} to verify properties in OpenDP; however, that work
focuses on non-probabilistic functional correctness properties of parts of the library, in contrast to the probabilistic samplers we have verified.

\paragraph{Prior Verifications of Similar Case Studies}
The family of discrete Gaussian samplers described in \Cref{sec:cks}
has also been verified in SampCert~\cite{sampcert}.
SampCert is written in a probabilistic-monadic DSL, called SLang, embedded in Lean.
Sampler correctness is established by showing that the monadic denotation of the program is the intended distribution.
Executable samplers are produced by extracting or compiling this DSL code.
We instead reason about the executable Rust code syntactically through a program logic, using
error credits as a separation logic resource.
This also lets us verify interaction with Rust code that would fall outside a monadic DSL like SLang.

\citet{fldr_fm26} present a Hoare logic that they use to verify FDR and FLDR with a pencil-and-paper proof.
Their logic allows for stating \emph{distributional invariants}, which are invariants over the distribution of values stored in a variable.
This leads to elegant loop invariants and proofs for FDR and FLDR.
The lifting-based approach underlying Eris and hence \thetool{} does not support such distributional invariants.
Nevertheless, one can interpret the credit allocation function used in our proofs in \Cref{sec:fldr} as describing how the expected value transforms across iterations,
which \emph{does} indirectly encode the distribution.
Moreover, the lifting-based approach allows us to reuse Verus's existing support for reasoning about the array-based representation of the FLDR tree.

\section{Results and Future Work}

We presented \thetool, a lightweight extension of Verus with probabilistic error
credits.
We demonstrated \thetool by verifying Rust samplers for the discrete Laplace and discrete Gaussian, the alias method,
and the fast loaded dice roller.  \thetool is able to take advantage of Verus's
existing SMT-backed automation and reasoning features, with soundness justified
through an Eris-based adaptation of VerusBelt.

For future work, it would be interesting to find a way to extend the soundness model to support prophecy variables.
While full prophecy variables are not compatible with Eris, Verus only requires a limited form of prophecies.
Verus imposes restrictions on prophecies to prevent inconsistencies with other Verus features, and these
restrictions might make them compatible with error credits.
Additionally, \thetool could be extended to reason about concurrent systems~\cite{coneris}
or to support relational properties of probabilistic programs, such as differential privacy~\cite{clutch-dp, lightdp}.

\begin{acks}
We thank the Verus community for their support, especially 
Baltasar Dinis, Travis Hance, Chris Hawblitzel, Bryan Parno, and Daniel Schoepe.
We also thank Alejandro Aguirre, Markus de Medeiros, 
Lars Birkedal, Simon Oddershede Gregersen, Philipp Haselwarter, Kwing Hei Li, 
and Puming Liu for their helpful discussions.
The first author especially thanks Markus for his help on demystifying the 
Fast Geometric Exponential sampler.
This work was supported in part by the \grantsponsor{NSF}{National Science Foundation}{} under Grant No.~\grantnum{NSF}{2338317}.
Any opinions, findings, and conclusions or recommendations expressed in this material are those
of the authors and do not necessarily reflect the views of these funding agencies.
\end{acks}

\bibliography{refs}

\ifbool{fullversion}{
\pagebreak
\appendix

\section{Convergence of the Random-Walk Fail Probability}
\label{sec:rw-convergence}

Recall the credit allocation for the 1-dimensional random walk,
\[
  \mathtt{fail\_prob}(s, p) \;=\; 1 - \Pr[\text{walk from } p \text{ reaches } 0 \text{ within } s \text{ steps}],
\]
which satisfies $\mathtt{fail\_prob}(s, 0) = 0$, $\mathtt{fail\_prob}(0, p) = 1$ for $p > 0$, and
$\mathtt{fail\_prob}(s, p) = \tfrac{1}{2}\bigl(\mathtt{fail\_prob}(s{-}1, p{-}1) + \mathtt{fail\_prob}(s{-}1, p{+}1)\bigr)$.
To justify picking a sufficient fuel we must show
\[
  \forall \delta > 0.\;\exists s.\; \mathtt{fail\_prob}(s, \mathtt{pos}) < \delta.
\]
We prove this by (i) showing
$\mathtt{fail\_prob}(\cdot,\, p)$ is non-increasing and bounded in $[0,1]$, so by monotone
convergence it has a limit $\mathtt{fail\_limit}(p)$; (ii) taking limits in the recurrence
to obtain the equation
\[
  \mathtt{fail\_limit}(p) \;=\; \tfrac{1}{2}\bigl(\mathtt{fail\_limit}(p{-}1) + \mathtt{fail\_limit}(p{+}1)\bigr),
  \qquad \mathtt{fail\_limit}(0) = 0,
\]
whose general solution forces
\[
  \mathtt{fail\_limit}(p) \;=\; p \cdot \mathtt{fail\_limit}(1);
\]
and (iii) ruling out $\mathtt{fail\_limit}(1) > 0$,
which would give $\mathtt{fail\_limit}(p) > 1$ for some $p$, contradicting the bound.
Hence $\mathtt{fail\_limit}(p) = 0$ for every $p$, so $\mathtt{fail\_prob}(\cdot, \mathtt{pos})$
converges to $0$ and drops below any $\delta > 0$ at some finite $s$.

\section{Invariant Preservation for the Negative-Exponential Bernoulli Sampler}
\label{sec:bern-exp1-invariant}

Recall the loop invariant for \texttt{sample\_bernoulli\_exp1} from \Cref{sec:cks-bern-exp1}: on entry
to iteration $k$ the held credit $\err_k$ satisfies
$\err_k \ge p_k\,\Err(\texttt{true}) + (1-p_k)\,\Err(\texttt{false})$, where the return probabilities
$p_k$ obey $p_k = \tfrac{x}{k}\,p_{k+1} + (1-\tfrac{x}{k})\,[k\text{ odd}]$. On the \texttt{true}
branch of iteration $k$ the credit carried into the next iteration is
$\err_{k+1} = \tfrac{k}{x}\,\err_k - (\tfrac{k}{x}-1)\,\Err([k\text{ odd}])$, and we check that the
invariant still holds at step $k+1$:
\[
\begin{aligned}
  \err_{k+1}
  &= \tfrac{k}{x}\,\err_k - \bigl(\tfrac{k}{x}-1\bigr)\Err([k\text{ odd}])\\
  &\ge \tfrac{k}{x}\bigl(p_k\,\Err(\texttt{true}) + (1-p_k)\,\Err(\texttt{false})\bigr)
        - \bigl(\tfrac{k}{x}-1\bigr)\Err([k\text{ odd}])\\
  &= p_{k+1}\,\Err(\texttt{true}) + (1-p_{k+1})\,\Err(\texttt{false}),
\end{aligned}
\]
using $\Err([k\text{ odd}]) = [k\text{ odd}]\,\Err(\texttt{true}) + (1-[k\text{ odd}])\,\Err(\texttt{false})$
and the recurrence rearranged as
$\tfrac{k}{x}\,p_k - \bigl(\tfrac{k}{x}-1\bigr)[k\text{ odd}] = p_{k+1}$.
The invariant is thus preserved, and since $p_{k+1}\in[0,1]$ and $\Err \ge 0$ the resulting
$\err_{k+1}$ is hence a valid credit.

\section{The Taylor-Tail Axiom for $e^{-x}$}
\label{sec:taylor-axiom}

Verus's SMT backend has no theory of Euler's number $e$, nor does it have completeness of reals.
\thetool\ exposes $e^{(-)}$ as an uninterpreted function and axiomatizes only the facts it needs. 
The single nontrivial axiom is the alternating-series bound on the Taylor partial sums of
$e^{-x}$, which the negative-exponential sampler of \autoref{sec:cks-bern-exp1} uses to show its
conditional probabilities $p_k$ lie in $[0,1]$. We discharge it separately in Lean, so the
Verus-level development depends only on this small, clearly delimited interface.

\begin{lstlisting}[language=Verus,style=VerusLineNos]
pub uninterp spec fn exp(x: real) -> real;                    // e^x, uninterpreted

pub open spec fn exp_taylor_term(x: real, k: nat) -> real {
    pow(-x, k) / factorial(k)                                 // (-x)^k / k!
}
pub open spec fn exp_taylor_seq(x: real) -> spec_fn(nat) -> real {
    |k: nat| exp_taylor_term(x, k)
}

// T_n = partial_sum(exp_taylor_seq(x), n) = sum_{k<n} (-x)^k / k!
#[verifier::external_body]
pub proof fn axiom_exp_taylor_bounds(x: real, n: nat)
    requires 0real < x <= 1real, n >= 1,
    ensures
        0real <= partial_sum(exp_taylor_seq(x), n),           // T_n in [0,1]
        partial_sum(exp_taylor_seq(x), n) <= 1real,
        n %
        n %
{}
\end{lstlisting}

\section{The Fast Geometric Sampler}
\label{sec:fast-geometric}

For $x = n/d$ in lowest terms, \texttt{sample\_geometric\_exp\_fast} draws $u$ from the
exponential-rejection sampler, $\Pr[u] = e^{-u/d}/N$ with $N = \sum_{u=0}^{d-1} e^{-u/d}$, and
an independent $v \sim \mathrm{Geom}(1-e^{-1})$ with $\Pr[v] = e^{-v}(1-e^{-1})$, and returns
$\lfloor (u + d\,v)/n\rfloor$. Working backwards from \rref{EPT}, the slow geometric sampler is handed
$g(u,v) = \Err\big((u + d\,v)/n\big)$, forcing the rejection sampler's allocation
\[
  f(u) \;=\; \expect[v\sim\mathrm{Geom}(1-e^{-1})]{g(u,v)}
       \;=\; \sum_{v=0}^{\infty} e^{-v}(1-e^{-1})\,\Err\big((u + d\,v)/n\big).
\]
The precondition $\err = \sum_{r=0}^{\infty}(e^{-n/d})^r(1-e^{-n/d})\,\Err(r)$ then covers the rejection
sampler's credit, using two Euclidean-division bijections, $k := u + d\,v$ over
$\{0,\dots,d-1\}\times\mathbb{N}$ and $k = n\,r + i$ over $\mathbb{N}\times\{0,\dots,n-1\}$:
\[
\begin{aligned}
  \expect[u\sim\mathrm{rejection}]{f(u)}
  &= \tfrac{1}{N}\sum_{u=0}^{d-1} e^{-u/d}\, f(u)\\
  &= \tfrac{1-e^{-1}}{N}\sum_{u=0}^{d-1}\sum_{v\in\mathbb{N}} e^{-(u+ dv)/d}\,\Err\big((u + d\,v)/n\big)
     && \text{(unfold } f)\\
  &= \tfrac{1-e^{-1}}{N}\sum_{k\in\mathbb{N}} e^{-k/d}\, \Err(k/n)
     && (k = u + d\,v)\\
  &= \tfrac{1-e^{-1}}{N}\sum_{r\in\mathbb{N}}\sum_{i<n} e^{-(nr+i)/d}\, \Err(r)
     && (k = n\,r + i)\\
  &= \tfrac{1-e^{-1}}{N}\,\tfrac{1-e^{-n/d}}{1-e^{-1/d}}\sum_{r\in\mathbb{N}} (e^{-n/d})^r\,\Err(r)\\
  &= \sum_{r\in\mathbb{N}} (e^{-n/d})^r (1 - e^{-n/d})\,\Err(r)
     && \bigl(N = \tfrac{1-e^{-1}}{1-e^{-1/d}}\bigr).
\end{aligned}
\]

\section{Credit Split for the Discrete Laplace Sampler}
\label{sec:discrete-laplace-credit}

The discrete Laplace sampler of \Cref{sec:discrete-laplace} flips a fair sign and then, inside the
chosen branch, draws the magnitude $k$ with \texttt{sample\_geometric\_exp\_fast}, a
$\mathrm{Geom}(1-p)$ sampler (\Cref{sec:geometric-bernoulli}). We hand that magnitude draw the
allocation $k \mapsto \Err(k)$ in the positive branch and $k \mapsto \Err_{\!-}(k)$ in the negative
branch, where $\Err_{\!-}(k) = \Err(-k)$ for $k \ge 1$ and $\Err_{\!-}(0) = \err$. 
By that sampler's \rref{EPT}
precondition, the credit each branch must supply is the 
$\mathrm{Geom}(1-p)$-average of its allocation,
so backward execution through the sign flip yields the two branch budgets
\[
  \err_{+} = \expect[k\sim\mathrm{Geom}(1-p)]{\Err(k)} = \sum_{k} p^k(1-p)\,\Err(k),
  \qquad
  \err_{-} = \sum_{k} p^k(1-p)\,\Err_{\!-}(k).
\]
The sign flip's own \rref{EPT} precondition then requires its average $\tfrac12(\err_{+} + \err_{-})$
to fit within $\err$. Unfolding the two budgets and splitting off the $k=0$ term of each,
\[
\begin{aligned}
  \err_{+} + \err_{-}
  &= (1-p)\,\err + (1-p)\,\Err(0) + \sum_{k\ge 1} p^k(1-p)\bigl(\Err(k) + \Err(-k)\bigr)\\
  &= (1-p)\,\err + (1+p)\Bigl(\mu_L(0)\,\Err(0) + \sum_{k\ge 1}\bigl(\mu_L(k)\,\Err(k) + \mu_L(-k)\,\Err(-k)\bigr)\Bigr)\\
  &= (1-p)\,\err + (1+p)\sum_{x\in\mathbb{Z}} \mu_L(x)\,\Err(x)
   \;\le\; (1-p)\,\err + (1+p)\,\err \;=\; 2\,\err,
\end{aligned}
\]
so $\tfrac12(\err_{+} + \err_{-}) \le \err$ and the sign flip's precondition is met.

\section{The Discrete Gaussian Sampler}
\label{sec:discrete-gaussian-appendix}

The discrete Gaussian $\mathcal{N}_{\mathbb{Z}}(0,\sigma^2)$ has $\mathrm{pmf}(x) = e^{-x^2/2\sigma^2}/Z$
with $Z = \sum_{y\in\mathbb{Z}} e^{-y^2/2\sigma^2}$. As noted in \Cref{sec:discrete-gaussian}, it is
sampled by rejection against a discrete-Laplace proposal: with $t = \lfloor\sigma\rfloor + 1$, we draw
$y \sim \mathrm{Lap}(0,t)$ and accept with probability $C(y) = e^{-\mathrm{bias}(y)}$, where
$\mathrm{bias}(y) = (|y| - \sigma^2/t)^2/(2\sigma^2)$.
\begin{lstlisting}[language=Verus,style=VerusLineNos]
pub fn sample_discrete_gaussian(sigma: &RBig) -> IBig {
    let t: RBig = (sigma.floor() + 1).into();
    let sigma2 = sigma * sigma;
    loop {
        let y = sample_discrete_laplace(&t);
        let bias = (y.abs() - &sigma2 / &t).square() / (2 * &sigma2);
        if sample_bernoulli_exp(&bias) {
            return y;
        }
    }
}
\end{lstlisting}

Writing $p = e^{-1/t}$, the discrete-Laplace proposal has pmf $\mu_L(y) = \tfrac{1-p}{1+p}\,e^{-|y|/t}$.
The key identity is that the proposal weight times the acceptance probability is a multiple of the
target: since $|y|/t + \mathrm{bias}(y) = y^2/2\sigma^2 + \sigma^2/2t^2$,
\[
  \mu_L(y)\,C(y)
  = \tfrac{1-p}{1+p}\,e^{-|y|/t}\,e^{-\mathrm{bias}(y)}
  = \kappa\, e^{-y^2/2\sigma^2}
  = a\,\mathrm{pmf}(y),
  \qquad
  \kappa = \tfrac{1-p}{1+p}\,e^{-\sigma^2/2t^2},
  \quad
  a = \kappa Z,
\]

Summing over $y$, one iteration accepts
with probability $a = \sum_y \mu_L(y)\,C(y) = \kappa Z$, and conditioned on accepting it returns $y$ with
probability $\kappa\,e^{-y^2/2\sigma^2}/a = \mathrm{pmf}(y)$.

As in all rejection samplers, the loop maintains $\err \ge \sum_{x\in\mathbb{Z}} \mathrm{pmf}(x)\,\Err(x)$.
The acceptance flip $\mathrm{Bern}(C(y))$ pays out $\Err(y)$ on \texttt{true} (accept and return $y$) and
carries $\err$ back on \texttt{false} (reject and restart), so backward execution through the
$\mathrm{Lap}(0,t)$ proposal draw forces the allocation
$g(y) = C(y)\,\Err(y) + (1 - C(y))\,\err$. Discharging its \rref{EPT} precondition,
$\sum_y \mu_L(y)\,g(y) \le \err$, unfolds via the factorization
$\mu_L(y)\,C(y) = a\,\mathrm{pmf}(y)$ above as
\[
\begin{aligned}
  \sum_y \mu_L(y)\,g(y)
  &= \sum_y \mu_L(y)\,C(y)\,\Err(y) + \err\sum_y \mu_L(y)\bigl(1 - C(y)\bigr)\\
  &= \sum_y a\,\mathrm{pmf}(y)\,\Err(y) + \err\Bigl(\textstyle\sum_y \mu_L(y) - \sum_y \mu_L(y)\,C(y)\Bigr)
     && \bigl(\mu_L(y)\,C(y) = a\,\mathrm{pmf}(y)\bigr)\\
  &= a\sum_{x\in\mathbb{Z}} \mathrm{pmf}(x)\,\Err(x) + (1 - a)\,\err
     && \bigl(\textstyle\sum_y \mu_L(y) = 1,\ \sum_y \mu_L(y)\,C(y) = a\bigr)\\
  &\le a\,\err + (1 - a)\,\err \;=\; \err,
\end{aligned}
\]
the inequality by the precondition $\sum_x \mathrm{pmf}(x)\,\Err(x) \le \err$. So the held credit
$\err$ covers the proposal draw: on acceptance the loop returns $y$ holding $\Err(y)$, and on
rejection it carries $\err$ back to restart, exactly as in the discrete-Laplace loop.

\section{Mathematical Bound on FDR Credit Allocation}
\label{sec:fdr-bound}

We prove the bound by summing all possible outcomes for $c \in [0,v)$:  $S(v,k) = \sum_{c<v}\mathrm{fdr}_f(v,c,k)$
and establishing the \emph{uniform-in-$v$} bound:
\[
  \forall v.\quad S(v,k) \;\le\; v\cdot\mathrm{avg}(n,\mathcal{E}).
\]
We prove this by induction on fuel $k$. Unfolding $\mathrm{fdr}_f$ and reindexing the resulting pair sum,
\[
  S(v,k)
  \;=\; \sum_{c<v}\tfrac12\bigl(\mathrm{fdr}_h(2v,\,2c,\,k{-}1) + \mathrm{fdr}_h(2v,\,2c{+}1,\,k{-}1)\bigr)
  \;=\; \tfrac12\sum_{c<2v}\mathrm{fdr}_h(2v,\,c,\,k{-}1),
\]
and, when $2v \ge n$, a threshold split at $c = n$ separates the accept terms from the reject terms,
\[
\begin{aligned}
  S(v,k) &= \tfrac12\sum_{c<2v}\mathrm{fdr}_h(2v,\,c,\,k{-}1)\\[4pt]
  &= \tfrac12\Bigl(\,\underbrace{\sum_{c<n}\mathcal{E}(c)}_{\text{accept}}
        \;+\; \underbrace{\sum_{n\le c<2v}\mathrm{fdr}_f(2v{-}n,\,c{-}n,\,k{-}1)}_{\text{reject, restart}}\,\Bigr)\\[4pt]
  &= \tfrac12\bigl(\,\textstyle\sum_{i<n}\mathcal{E}(i) + S(2v{-}n,\,k{-}1)\,\bigr)\\[2pt]
  &\le \tfrac12\bigl(\,\textstyle\sum_{i<n}\mathcal{E}(i) + (2v{-}n)\cdot\mathrm{avg}(n,\mathcal{E})\,\bigr)
       &&\text{(IH)}\\[2pt]
  &= \tfrac12\bigl(\,n + (2v{-}n)\,\bigr)\cdot\mathrm{avg}(n,\mathcal{E})
   \;=\; v\cdot\mathrm{avg}(n,\mathcal{E}),
       &&\bigl(\textstyle\sum_{i<n}\mathcal{E}(i) = n\cdot\mathrm{avg}(n,\mathcal{E})\bigr)
\end{aligned}
\]
where the second equality reindexes $c \mapsto c - n$ over $[n,2v)$, and the inequality applies the
inductive hypothesis at fuel $k{-}1$. The sub-threshold case $2v < n$ is
similar but with no accept terms. The base case $k = 0$ is immediate, since
$\mathrm{fdr}_f(\cdot,0) = 0$ makes $S(v,0) = 0$.
Finally, instantiating the uniform bound at the loop's start level $v = 1$ gives the claim: the sum
$S(1,k) = \sum_{c<1}\mathrm{fdr}_f(1,c,k)$ has the single term $\mathrm{fdr}_f(1,0,k)$, so
$\mathrm{fdr}_f(1,0,k) \le \mathrm{avg}(n,\mathcal{E})$.

\paragraph{The loaded case (FLDR)}
The verification uses the analogous pair of mutually recursive credit allocations, now following the
DDG tree's level/position structure rather than the doubling window. Here $\mathrm{fldr}_f(c,d,k)$ is
the conditional expectation $\expect{\mathcal{E}(\mathit{out}) \mid (c,d)}$ over the next $k$ coin flips
from the node at depth $c$ and position $d$, and $\mathrm{fldr}_g$ resolves a node after the flip:
\[
\begin{aligned}
  \mathrm{fldr}_f(c,d,0) ={}& 0                          &\qquad\quad& &\qquad\quad& \text{(ran out of fuel $k$)}\\
  \mathrm{fldr}_f(c,d,k) ={}& \rlap{$\tfrac12\bigl(\mathrm{fldr}_g(c{+}1,\,2d,\,k{-}1)
                              + \mathrm{fldr}_g(c{+}1,\,2d{+}1,\,k{-}1)\bigr)$} &\qquad\quad& &\qquad\quad&\\[4pt]
  \mathrm{fldr}_g(c,d,k) ={}& \mathcal{E}(\mathit{lab}[c][d])   &\qquad\quad& \text{if } d < h[c],\ \mathit{lab}[c][d] < n &\qquad\quad& \text{(accept)}\\
                         {}& \mathrm{fldr}_f(0,\,0,\,k)         &\qquad\quad& \text{if } d < h[c],\ \mathit{lab}[c][d] = n &\qquad\quad& \text{(reject, restart)}\\
                         {}& \mathrm{fldr}_f(c,\,d{-}h[c],\,k)  &\qquad\quad& \text{if } d \ge h[c]                        &\qquad\quad& \text{(internal, descend)}
\end{aligned}
\]
Mimicking the program, $\mathrm{fldr}_g$ pays out $\mathcal{E}(\mathit{lab}[c][d])$ on a real-outcome
leaf, restarts at the root node $(0,0)$ on a reject leaf ($\mathit{lab}[c][d] = n$), and otherwise
renumbers $d \mapsto d - h[c]$ to descend into the internal nodes. The termination allocation
$\mathrm{fldr\_fail}_f$ again mirrors this recursion, with the accepting payout replaced by $0$ and
the out-of-fuel base case set to $1$.

The fast loaded dice roller bound is proved the same way: we show the claim
$\mathrm{fldr}_f(0,0,k) \le T$ ($T := \textstyle\sum_{i<n}\tfrac{a_i}{m}\,\mathcal{E}(i)$) 
by induction on the fuel $k$. 
Unfolding $\mathrm{fldr}_f(0,0,k)$: each leaf $(c,d)$ is reached with probability
$2^{-c}$, 
an accept leaf contributes $\mathcal{E}(\mathit{lab}[c][d])$, 
each reject leaf restarts at the root with the remaining fuel $k - c$:
\[
\begin{aligned}
  \mathrm{fldr}_f(0,0,k)
  &= \sum_{\text{accept}} 2^{-c}\,\mathcal{E}(\mathit{lab})
     \;+\; \sum_{\text{reject}} 2^{-c}\,\mathrm{fldr}_f(0,0,k-c)\\[2pt]
  &\le \sum_{\text{accept}} 2^{-c}\,\mathcal{E}(\mathit{lab})
     \;+\; T\sum_{\text{reject}} 2^{-c}
     &&\text{(IH)}\\[2pt]
  &= \tfrac{m}{2^{\mathit{depth}}}\,T \;+\; \Bigl(1 - \tfrac{m}{2^{\mathit{depth}}}\Bigr)\,T \;=\; T,
     &&\text{(leaf-sum identity)}\\[4pt]
  &\hspace{-2.2em}\text{where } \mathit{depth} = \lceil \log_2 m \rceil
\end{aligned}
\]
The regrouping is the DDG leaf-sum identity: collecting the accept leaves by label gives
$\sum_{\text{accept}} 2^{-c}\mathcal{E}(\mathit{lab}) = \sum_{i<n}\tfrac{a_i}{2^{\mathit{depth}}}\mathcal{E}(i)
= \tfrac{m}{2^{\mathit{depth}}}\,T$, while the reject leaves carry
$\sum_{\text{reject}} 2^{-c} = 1 - m/2^{\mathit{depth}}$. The induction is well founded because
every leaf has depth ${\ge}\,1$.

}{}

\end{document}